\documentclass[pre,aps,floatfix,twocolumn,reprint]{revtex4-2}%showpacs
\usepackage[utf8]{inputenc}
\usepackage[T1]{fontenc}

\usepackage{graphicx}% Include figure files
\usepackage{dcolumn}% Align table columns on decimal point
\usepackage{bm}% bold math
\usepackage{amssymb}
\usepackage{pifont}
\usepackage{amsmath}
\usepackage{amsthm}
\usepackage{times}
\usepackage{epstopdf}
\usepackage{latexsym}
\usepackage{keyval}
\usepackage{ifthen}
\usepackage{moreverb}
\usepackage{color}

\usepackage[hidelinks,unicode=true]{hyperref}

\usepackage{lipsum}
\usepackage{autonum}

\newcommand{\KY}{ \mathcal{D}_\text{KY} }

\DeclareMathOperator{\acos}{acos}
\DeclareMathOperator{\Real}{Re}

\DeclareMathOperator{\Imag}{Im}

\DeclareMathOperator{\sign}{sign
}

\renewcommand{\selectlanguage}[1]{} % To solve the bibtex issue

\graphicspath{{./figures/}}

\begin{document}

\title{Symmetry breaking and high-dimensional chaos in sparse random networks of exact firing rate models}
\author{Pau Clusella}
\affiliation{Departament de Matemàtiques, Universitat Politècnica de Catalunya, Manresa, Spain}

\begin{abstract}
Exact firing rate models, also known as next-generation neural mass models (NG-NMMs), provide a rigorous description
of the dynamics of neural populations. 
While in its simplest form a single population only displays fixed-point activity, multi-population models may display a range
of different behaviors.
In this work, we study the dynamics of all-excitatory or all-inhibitory NG-NMMs coupled through sparse random networks with row-normalized network topology.
Linear stability analysis of the homogeneous states of the system, representing asynchronous neural activity, provides
a dispersion relation linking the emergence of spatiotemporal dynamics to the spectra of the connectivity matrix.
Using bounds from random matrix theory, we identify the parameter regions where instabilities occur. 
In undirected networks, only inhibitory systems produce heterogeneous stationary patterns, corresponding to a winner-takes-all mechanism. 
In directed networks, exotic rhythmic states with high frequencies emerge in both, excitatory and inhibitory systems. 
Numerical simulations reveal that these hectic oscillatory states correspond to high-dimensional chaos with extensive properties.
\end{abstract}

\maketitle

\section{Introduction}

Exact firing rate models, also known as Next-Generation Neural Mass Models (NG-NMMs), represent a major breakthrough in understanding
the collective behavior of populations of neurons \cite{Montbrio2015,Luke2013,Coombes2019,bick2020,castaldo2026}.
By means of an exact mean-field theory, NG-NMMs provide a low-dimensional
system for the dynamics of globally coupled quadratic integrate-and-fire (QIF) neurons,
thus rigorously bridging single-cell dynamics with mesoscale activity. 
This framework has subsequently been extended to increasingly complex neuronal dynamics through both exact and approximate mean-field theories\cite{Pietras2019,Goldobin2021,chen2022,clusella_exact_2024,PP22,pietras2025,pazo2025,cestnik2026}.
It has also revealed different forms of collective behavior in one and two population models, including rhythmic activity,
multistability, and symmetry breaking \cite{Pazo2016,Devalle2017,Devalle2018,diVolo2018,Ratas2019,Dumont2019,reyner-parra2021,Segneri2020,Bi2020,Clusella2022,pyragas2023,pietras2024,mayora-cebollero2025}.

Recent studies have explored the dynamics emerging from coupling NG-NMMs
through whole-brain architectures \cite{Gerster2021,Perl2023,clusella2023,Forrester2024,delicado-moll2026}.
In this context, each network node represents a functional brain region whose dynamics are described by a NG-NMM.
In particular, in \cite{clusella2023,delicado-moll2026} it has been shown that complex spatiotemporal behavior, including traveling waves and high-dimensional chaotic behavior, emerges from transverse instabilities of a homogeneous state.
However, these models rely on the interplay between excitatory and inhibitory populations within each network node. 
Therefore, even an isolated brain region exhibits a rich dynamical repertoire, including limit cycles, bistability, and low-dimensional chaos.
Consequently, it remains unclear whether such intrinsic single-node complexity is necessary for the emergence of spatiotemporal behavior.
Moreover, because these studies focus on fixed, empirically derived connectomes, the role of network topology in shaping the emergence of novel dynamical states also remains poorly understood.

In this work, we take a significant leap forward in understanding the behavior of NG-NMM interacting through complex topologies. Specifically, we address a simple yet intricate question: what are the collective dynamics of random networks of NG-NMMs where each node represents a single QIF population? We focus on purely excitatory or purely inhibitory populations coupled through instantaneous interactions. In this regime, isolated nodes are dynamically trivial: the corresponding NG-NMM converges to a stable asynchronous state \cite{Montbrio2015}.

To analyze the stability of these states, we impose a row-normalization constraint on the connectivity matrix, which guarantees the existence of a homogeneous manifold. This allows the stability problem to be formulated in a manner analogous to Turing instabilities in networked systems \cite{Nakao2010,Asllani2014}. In particular, we derive a dispersion relation linking perturbation growth rates to the eigenvalue spectrum of the connectivity matrix. Combining this relation with analytical results from random matrix theory yields explicit conditions for the onset of instabilities in terms of network structure and coupling parameters.

Overall, our analysis reveals two qualitatively distinct instability mechanisms: 
In undirected networks, instabilities emerge only for inhibitory populations,
leading to the formation of steady patterns via a winner-takes-all mechanism. 
Instead, in directed networks we observe the emergence of irregular oscillatory states
regardless of the coupling sign. 
Detailed numerical analysis of these dynamics via Lyapunov exponents shows that this is a case
of high-dimensional and extensive resonant chaos \cite{muscinelli2019}. 
Remarkably, we show that, in all cases, network sparsity is a fundamental ingredient for the emergence of these complex states.

\section{The model and stability analysis}

\subsection{Networks of exact Firing Rate Equations}

A system composed of a formally infinite number of globally coupled quadratic integrate-and-fire (QIF) neurons admits an exact
low-dimensional reduction \cite{Montbrio2015}.
In its simplest form, the resulting neural mass model reads
\begin{align}
    \tau \dot r &= \frac{\Delta}{\pi \tau} + 2rv\\
    \tau \dot v &= \eta + v^2 - (\pi \tau r)^2 + I(t).
\end{align}
where $r$ (kHz) is the mean firing rate of the population, $v$ is the mean membrane potential, 
$\eta$ is the mean constant current affecting all neurons equally,
$\Delta$ accounts for either the noise or heterogeneity of the input currents given by a Cauchy distribution \cite{clusella_exact_2024},
$I(t)$ include any other common inputs,
and $\tau $ (ms) is a time scale parameter.

Here we consider a network of $N$ \emph{populations} of QIF neurons interacting through instantaneous synaptic dynamics:
\begin{align}\label{eq:system1}
    \tau \dot r_i &= \frac{\Delta}{\pi \tau} + 2r_iv_i\\
    \tau \dot v_i &= \eta + v_i^2 - (\pi \tau r_i)^2 + J\tau \sum_{j=1}^N c_{ij} r_j
\end{align}
for $i=1,\dots,N$.
Here $J$ is a global coupling strength
and $C=(c_{ij})$ is the connectivity matrix of the network, which satisfies the row-normalization condition
\begin{equation}\label{eq:normalization}
    \sum_{j=1}^N c_{ij}=1\quad\forall i=1,\dots,N.
\end{equation}
To keep the model as simple as possible, we assume that neural populations are either all excitatory ($J>0$) or all inhibitory ($J<0$),
with $c_{ij}\geq 0$ for all $i,j=1,\dots,N$.
Due to their role in Markov processes, matrices with these properties are known as \emph{right-stochastic matrices}.

System \eqref{eq:system1} admits a parameter reduction through variable and parameter rescaling.
Therefore, we fix $\Delta=1$ and $\tau=10$ ms for the rest of the paper without loss of generality \footnote{Consider the new dependent variables $R_i=\tau r_i/\Delta^{1/2}$ and $V_i=v_i/\Delta^{1/2}$, and the time rescaling $T = t \Delta^{1/2}/\tau$.
Then, the resulting system for $dR_i/dT$ and $dV_i/dt$ depends only on the new parameters $\tilde \eta = \eta/\Delta$ and $\tilde J = J/\Delta^{1/2}$.}. All variables and parameters of the system are dimensionless, except for the time-related units $t$ and $\tau$, which are
given in milliseconds, and the firing rates $r_i$ for $i=1,\dots,N$, which are given in kilohertz \cite{Devalle2017,Clusella2022} 

\subsection{Homogeneous dynamics}

\begin{figure}[t]
  \centerline{\includegraphics[width=0.5\textwidth]{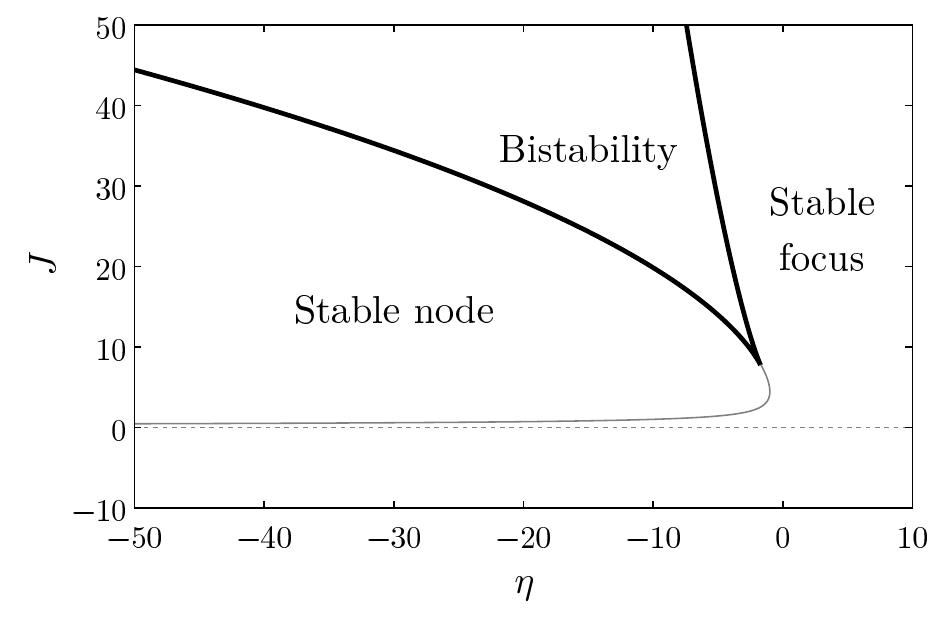}}
        \caption{Two-parameter bifurcation diagram of the homogeneous system \eqref{eq:homogeneous}.
        Black curves indicate saddle-node bifurcations, which join in a cusp codimension-2 bifurcation. 
        Grey thin curve indicates the focus-node boundary of the stable fixed point. 
        }
  \label{fig:homogeneous}
\end{figure}

Due to the row-normalization condition \eqref{eq:normalization}, Eq. \eqref{eq:system1} accepts a family
of homogeneous solutions.
Indeed, consider $r_i=r$ and $v_i=v$ for all $i=1,\dots,N$.
Substituting these expressions in system \eqref{eq:system1} we obtain
\begin{align}\label{eq:homogeneous}
    \tau \dot r &= \frac{\Delta}{\pi \tau} + 2rv\\
    \tau \dot v &= \eta + v^2 - (\pi \tau r)^2 + J\tau r.
\end{align}
This model corresponds to a single population with recurrent coupling,
which was originally studied in \cite{Montbrio2015}. 
Next, we review the dynamics of this system, which are illustrated by the bifurcation diagram in Fig. \ref{fig:homogeneous}.

The fixed points of the homogeneous system \eqref{eq:homogeneous} are given by
\begin{align}\label{eq:fixedpoints}
\tau r_0 & = \Phi_\Delta (\eta+J\tau r_0)\quad\text{and}\\
v_0& = -\frac{\Delta}{2\pi\tau r_0}
\end{align}
where $\Phi_\Delta$ is the transfer function of the QIF population:
\begin{equation}
    \Phi_\Delta (I) = \frac{1}{\pi \sqrt{2}}\sqrt{I+\sqrt{I^2+\Delta^2}}.
\end{equation}
From these fixed point equations one can derive a relation between
the firing rate $r_0$ and the system parameters as:
\begin{equation}\label{eq:fp_relation}
    (\pi \tau r_0 )^4-J\pi^2 (\tau r_0)^3 - \eta(\pi \tau r_0 )^2 = \frac{\Delta^2}{4}.
\end{equation}

Analysis of the fixed points given by Eqs. \eqref{eq:fixedpoints} and their stability shows that,
in excitatory networks ($J>0$), the parameter space is divided in two regions, depicted in Fig~\ref{fig:homogeneous}:
\begin{itemize}
    \item In a large region of the parameter space,
    including $\eta > 0$, a unique fixed point is the global attractor of the system.
    In most of this region, the fixed point is a focus, but for $\eta<0$ and low values of $J$ it becomes a node.
    Grey curve in Fig~\ref{fig:homogeneous} shows the node-focus transition.
    \item For $\eta < 0$, there exist a region of bistability between two fixed points bounded by two saddle-node bifurcations depicted by black curves in Fig~\ref{fig:homogeneous}.
    Within this region, the system may converge either to a persistent asynchronous state (equilibrium with higher firing rate)
    or to a low activity state. A third fixed point with intermediate values of $r_0$ remains unstable throughout. 
\end{itemize}
For inhibitory networks ($J<0$), the dynamics are even simpler, as there is a unique equilibrium, a stable focus, which is a global attractor of the system.

% % %%%%%%%%%%%%%%%%%%%%%%%%%%%%%%%%%%%%%%%%%%%%%%%%%%%%%%%%%%%%%%%%%%%%%%%%
% % %%%%%%%%%%%%%%%%%%%%%%%%%%%%%%%%%%%%%%%%%%%%%%%%%%%%%%%%%%%%%%%%%%%%%%%%

\subsection{Stability analysis and dispersion relation}

So far we have uncovered the dynamics of Eq. \eqref{eq:system1} under the condition of homogeneity.
In this section, we study the stability of the homogeneous fixed points $r_0$ and $v_0$ to nonuniform perturbations, i.e., 
perturbations \emph{transverse} to the homogeneous manifold given by Eq. \eqref{eq:homogeneous}.
This analysis is completely analogous to the study of Turing instabilities 
in complex networks \cite{Nakao2010,Asllani2014}.

We consider an arbitrary small perturbation of system \eqref{eq:system1}, $(\delta r_i,\delta v_i)$.
Linearizing, we obtain that
\begin{equation}\tau
    \begin{pmatrix}
        \dot \delta r_i\\ \dot \delta v_i
    \end{pmatrix}
    =
    \begin{pmatrix}
        2v_0 & 2 r_0\\
        -2(\pi \tau)^2 r_0 & 2v_0
    \end{pmatrix}
    +J\tau \sum_{j=1}^N c_{ij}
    \begin{pmatrix}
        0\\  \delta r_j
    \end{pmatrix}.
\end{equation}
Next, we perform the standard technique of decomposing the perturbation vector
on the basis given by the eigenvectors of $C$  (see Appendix \ref{ap:MSF}).
As a result,  the eigenvalues $\lambda_k$ controlling the
stability of the homogeneous state in Eqs. \eqref{eq:system1} are given by
the family of matrices
\begin{equation}
 \mathcal{J}_k = 
        \begin{pmatrix}
        2v_0 & 2 r_0\\
        -2(\pi \tau)^2 r_0 + J \tau \Lambda_k & 2v_0
    \end{pmatrix}\quad \text{for}\quad k=1,\dots,N;
\end{equation}
where $\Lambda_k$ are the eigenvalues of the connectivity matrix $C$.
In particular, the eigenvalues, $\lambda^{\pm}_k$ of $\mathcal{J}_k$ read
\begin{equation}\label{eq:msf}
    \lambda^\pm_k = 2v_0 \pm \sqrt{-2\tau r_0 (2\pi^2 \tau r_0 - J\Lambda_k)}.
\end{equation}
This equation constitutes
a dispersion relation between the structural eigenmodes $\Lambda_k$ and the growth rate $\lambda_k$ of a perturbation 
acting along the $k$-th eigenvector of $C$ (see Appendix \ref{ap:MSF} for full details).

Given the row-normalization condition Eq. \eqref{eq:normalization},
the Gershgorin circle theorem shows that $|\Lambda_k| \leq 1$ for all $k=1,\dots,N$ \cite{wilkinson1967}.
Moreover, there is always one unit eigenvalue, $\Lambda_1 = 1$, corresponding to a uniform eigenvector $\Psi_1 \propto (1,\dots,1)^T$.
Perturbations along this direction correspond to perturbations along the homogeneous manifold.
The emergence of inhomogeneous instabilities depends thus on the remaining eigenvalues $\Lambda_k$ for $k = 2,\dots,N$.
The distribution of these structural eigenvalues depend, in turn, on the specific properties of the topology chosen.

%-------------------------------------------------------------------------------------------------
%-------------------------------------------------------------------------------------------------
\section{Emergence of transverse instabilities in random networks}

\subsection{Spectra of right-stochastic random matrices}

% % %%%%%%%%%%%%%%%%%%%%%%%%%%%%%%%%%%%%%%%%%%%%%%%%%%%%%%%%%%%%%%%%%%%%%%%%
\begin{figure*}[t]
  \centerline{\includegraphics[width=\textwidth]{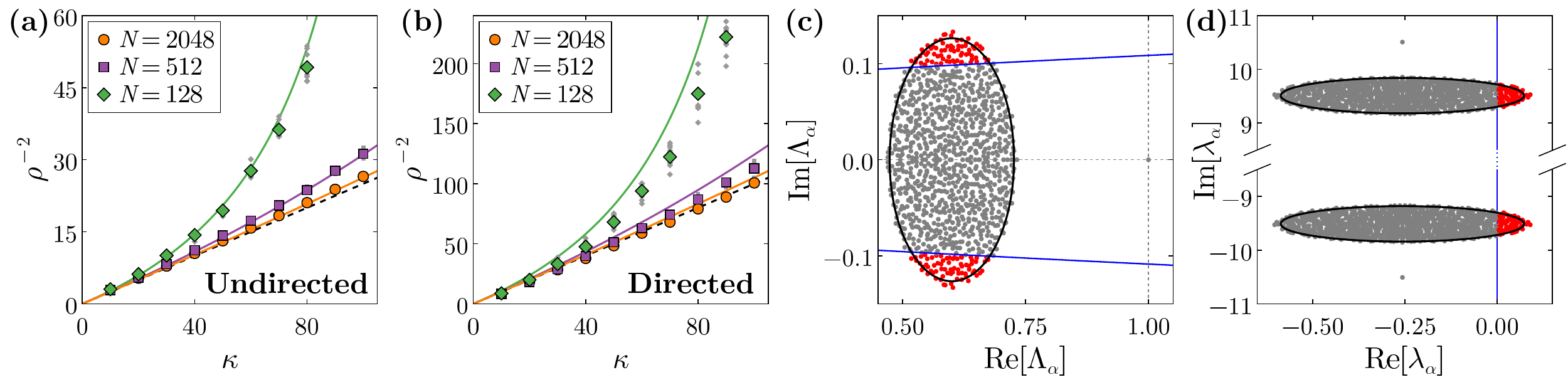}}
        \caption{Spectra of row-normalized connectivity matrices.
        (a,b) Scaling of the radius $\rho$ for the Wigner's semicircle (panel (a)) and Girko's disk (panel (b)) distributions of the bulk of eigenvalues of $C$.
        Continuous lines show the results corresponding to Eqs. \eqref{eq:spectra_undir} and \eqref{eq:spectra_undir} for panels (a) and (b) respectively.
        Symbols correspond to averages over 10 different networks for each combination of $\kappa$ and $N$, with grey dots showing
        the individual results. Black dashed lines indicate the asymptotic results $\rho^{-2} = \kappa/4$ (panel (a)) and $\rho^{-2} = \kappa$ (panel (b)). 
        (c) Spectra of a directed connectivity matrix $C$ obtained for $N=1024$, $\kappa=10$, and $\mu = 0.6$. 
        Red circles indicate the eigenvalues corresponding to unstable directions for $(\eta,J) =(40,-20)$,
        whereas gray circles correspond to stable directions. The blue continuous curves shows the instability boundary computed from Eq. \eqref{eq:complex_lambda}.
        (d) Jacobian eigenvalues corresponding to the spectra of $C$ depicted in panel (c), computed from Eq. \eqref{eq:complex_lambda}.
        }
  \label{fig:spectra}
\end{figure*}
% % %%%%%%%%%%%%%%%%%%%%%%%%%%%%%%%%%%%%%%%%%%%%%%%%%%%%%%%%%%%%%%%%%%%%%%%%

So far, we have not imposed any specific structure on $C$ except for the row-normalization constrain Eq. \eqref{eq:normalization}.
In the rest of the paper we study the case of neural mass models interacting through random (Erdős-Rényi) topologies constructed as follows:

Consider a network of $N$ interacting populations
connected through a (directed or undirected) Erdős-Rényi topology with fixed average degree $\kappa$.
For simplicity, we assume the network is fully connected, thus $\kappa >\log(N)$ \cite{erdos1959}. 
Let $A=(a_{ij})$ be the adjacency matrix of the network ($a_{ij}\in\{0,1\}$).
We introduce a new parameter $\mu\in[0,1]$ which establishes the degree of recurrent coupling
within each population, with $1-\mu$ determining the total amount
of coupling received from other populations.
Then, we define the coupling matrix $C=(c_{ij})$ as
\begin{equation}
    c_{ij} = \begin{cases}
    \mu & \text{if }\; i=j\\
    (1-\mu)\frac{a_{ij}}{d_i} & \text{if }\; i\neq j\\
    \end{cases}
\end{equation}
where $$d_i=\sum_{j=1}^N a_{ij}$$ is the (in)-degree of node $i$,
which we use to fulfill the row-normalization condition.
In matrix form, this can be expressed as:
\begin{equation}\label{eq:C}
    C = \mu I_N+ (1-\mu)D^{-1}A
\end{equation} 
where $D = (d_i\delta_{ij})$ is the diagonal matrix containing the in-degree of each node in the diagonal and
$\delta_{ij}$ is the Kronecker delta.

In order to analyze the stability of the system via \eqref{eq:msf} we need to study the eigenvalues of $C$, which is a right-stochastic random matrix. 
For our choice of $C$, it is enough to focus on the spectral properties of $D^{-1}A$,
since the eigenvalues of $C$ are
just a linear transformation of those.
Indeed, let $\Psi$ be an eigenvector of $D^{-1}A$ with associated eigenvalue $\tilde \Lambda$,
then $\Psi$ is also an eigenvector of $C$ with associated eigenvalue $\Lambda = \mu + (1-\mu) \tilde \Lambda$.
This result in turns shows that $C$ is diagonalizable if and only if so is $D^{-1}A$.

There is a vast ongoing mathematical literature research on the spectral properties of random matrices \cite{bai2010}.
Two main results are Wigner's semicircle law and Girko's circular law \cite{wigner1967,girko1985},
which provide the distribution for the eigenvalues of large symmetric and asymmetric random matrices respectively.
These classical theories do not directly apply to $D^{-1}A$ due to the row-normalization constrain.
However, equivalent results do exist for right-stochastic matrices \cite{bordenave2010,bordenave2012}.

Altogether, the spectra of $C$ as $N\to\infty$ has the two following properties: 
First, the eigenvalue $\Lambda_1=1$
associated to a uniform eigenvector is isolated \footnote{The multiplicity of this eigenvalue coincides with the number of closed strongly connected components in the graph.
Here we assume a strongly connected graph, i.e., $\kappa  > \log(N)$, thus the multiplicity of $\Lambda_1$ is one.}. 
Second, the other eigenvalues are distributed densely according to either a semicircle or a circular law:
\begin{enumerate}
    \item \textbf{Undirected networks:} In this case, $A$ is symmetric, thus the results in \cite{bordenave2010} apply. 
    Therefore $\Lambda_k$ are real and distributed according a Wigner's semicircle distribution centered at $\mu$
    and with radius 
    \begin{equation}\label{eq:spectra_undir}
     \rho = 2(1-\mu)\sqrt{\kappa^{-1} - N^{-1}}. %\fletxa  2(1-\mu)\sqrt{\kappa^{-1}}.
    \end{equation}
    \item \textbf{Directed networks:} In this case, following \cite{bordenave2012}, the bulk of the eigenvalues of $C$,
    are distributed uniformly within a disk in the complex plane centered at $\mu$ and with radius
    \begin{equation}\label{eq:spectra_dir}
     \rho = (1-\mu)\sqrt{\kappa^{-1}-N^{-1}}.% \fletxa  (1-\mu)\sqrt{\kappa^{-1}}.
    \end{equation}
\end{enumerate}
We remark that these results require $N\gg 1$ and no isolated nodes ($\kappa > \log(N)$). 
Given these constrains, the results are valid for both sparse ($\kappa \ll N$) and non-sparse ($\kappa \approx N$) topologies. 
In this later case, the radius of the semicircle and disk laws shrink, thus the bulk of eigenvalues concentrate around $\mu$.

To numerically test these results and their finite-size fluctuations, Figures \ref{fig:spectra}(a,b) 
show how the radius of numerically computed spectra compares to Eqs. \eqref{eq:spectra_undir} (panel (a)) and \eqref{eq:spectra_dir} (panel (b))
for networks of different sizes and average degree.
Numerically, we compute the radius as
$$\rho = \max_{k>1}|\Lambda_k-\mu|.$$
From Eqs. \eqref{eq:spectra_undir} and \eqref{eq:spectra_dir}, letting $\mu=0$ without loss of generality,
we have that
$$\rho^{-2} \approx q\frac{\kappa N}{N-\kappa}$$
where $q=0.25$ for undirected networks and $q=1$ for the directed case. 
Figures \ref{fig:spectra}(a,b) show that in both cases, the analytical relation (solid lines) agrees fairly well with the numerical results (symbols),
with indications of a clear convergence as the system size increases. The agreement
is better for the undirected case (panel (a)), with the directed networks displaying a consistent offset (panel (b)). 
Nonetheless, as $N$ grows, the asymptotic convergence to a semi-circle or a disk of radius $(q\kappa)^{-1/2}$
becomes evident. As an example, red and grey circles in figure \ref{fig:spectra}(c) show the eigenvalues of an undirected matrix $C$ for $N=1024$ 
and $\mu=0.6$,
with the black curve indicating the asymptotic domain given by Eq. \eqref{eq:spectra_dir}.

These boundaries for the distribution of the bulk of the spectra of $C$ allow
us to study the emergence of instabilities in random networks in a consistent manner.

%-------------------------------------------------------------------------------------------------
%-------------------------------------------------------------------------------------------------
\subsection{Bifurcation diagram}

% % %%%%%%%%%%%%%%%%%%%%%%%%%%%%%%%%%%%%%%%%%%%%%%%%%%%%%%%%%%%%%%%%%%%%%%%%
\begin{figure}[t]
  \centerline{\includegraphics[width=0.5\textwidth]{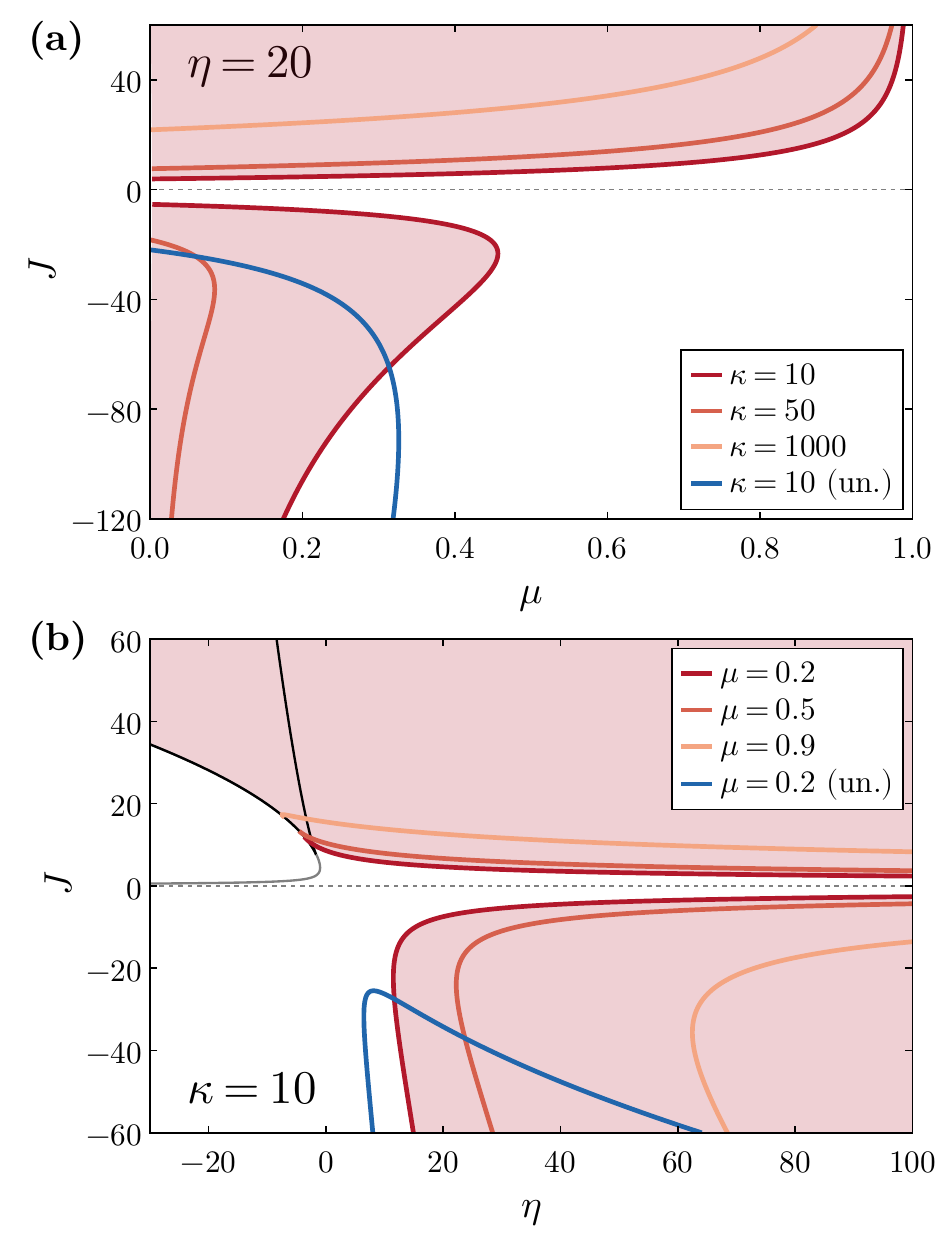}}
  \caption{Two-parameter bifurcation diagrams of the coupled system \eqref{eq:system1}.
        %(a) Diagram in the $(\mu,J)$ parameter space for fixed $\eta = 20$.
        (a) Red, salmon, and pink curves indicate the region of transverse instabilities for $\kappa=10$, $50$, and $1000$ respectively,
        with fixed $\eta=20$.
        The blue curve indicates the same result for an undirected network with $\kappa=10$.
        Red shaded region shows the region of transverse instabilities for a directed network with $\kappa = 10$.
        %(b) Diagram in $(\eta,J)$ parameter space for fixed $\kappa = 10$.
        (b) Red, salmon, and pink curves indicate the region of transverse instabilities for $\mu=0.2$, $0.5$, and $0.9$ respectively,
        with fixed $\kappa = 10$.
        The blue curve indicates the same result for an undirected network with $\mu=0.2$.
        Red shaded region shows the region of transverse instabilities for a directed network with $\mu = 0.2$.
        Black and gray curves correspond to bifurcations of the homogeneous dynamics as in Fig. \ref{fig:homogeneous}.
        }
  \label{fig:bifurcationdiagram}
\end{figure}
% % %%%%%%%%%%%%%%%%%%%%%%%%%%%%%%%%%%%%%%%%%%%%%%%%%%%%%%%%%%%%%%%%%%%%%%%%

Since the eigenvalue distribution of $C$ depends on whether the underlying network is directed or undirected,
we study these two cases separately. 

\subsubsection{Undirected networks}

In undirected networks $\Lambda_k\in\mathbb{R}$ for all $k=1,\dots,N$.
In excitatory populations ($J>0$), Eq. \eqref{eq:msf} shows that 
$$ \Real[\lambda_k] \leq \Real[\lambda_1 ]. $$
Since $\lambda_1$ corresponds to a perturbation along the homogeneous manifold,
 the homogeneous fixed points that are stable within Eqs. \eqref{eq:homogeneous} remain stable against nonhomogeneous perturbations. 
%\footnote{ Additional heterogeneous instabilities of the unstable equilibrium in the bistability region may arise.}.

In inhibitory networks ($J<0$), however, instabilities do arise.
Equation \eqref{eq:msf} shows that the first eigenmode to destabilize corresponds to the minimum $\Lambda_k$,
%which, following Eq.\eqref{eq:spectra_undir} with $N\gg 1$, is given by
which, following Eq.\eqref{eq:spectra_undir}, is given by
$$\Lambda_k \approx \mu - 2(1-\mu)\sqrt{\kappa^{-1}-N^{-1}}.$$
Inserting this expression into Eq.\eqref{eq:msf} and setting $\lambda_k^+=0$ we obtain the following condition
for the bifurcation:
\begin{align}\label{eq:transverse}
    J&\left(\mu-2(1-\mu)\sqrt{\kappa^{-1}-N^{-1}}\right)  \\&\qquad\quad= -\frac{\Delta^2}{2\pi^2 (\tau r_0)^3} - 2\pi^2 \tau r_0.
\end{align}
Since this instability corresponds to $\lambda_N\in\mathbb{R}$, no oscillatory components emerge in tangent space.
In other works, this corresponds to a classical Turing instability in a networked system \cite{Nakao2010}.

Equations \eqref{eq:fp_relation} and \eqref{eq:transverse} provide parametric curves on $r_0$
for the bifurcation boundary. 
Blue curves in Figures~\ref{fig:bifurcationdiagram} show the bifurcation in a $(\mu,J)$ and $(\eta,J)$ parameter
space, for $\kappa=10$ and $N\to \infty$.
The instability region occupies a large portion of the parameter space, but always requires low values of $\mu$.
Moreover, increasing network connectivity $\kappa$ causes this region to rapidly shrink and vanish (not shown) suggesting that low values of $\kappa$ are essential.
Indeed, a necessary condition for the bifurcation is that $\Lambda_k<0$ thus
$$\mu < \frac{2\sqrt{\kappa^{-1}-N^{-1}}}{1+2\sqrt{\kappa^{-1}-N^{-1}}}.$$
The right hand side of this inequality tends to $0$ as $\kappa \to N$, confirming thus
that network sparsity is a requirement for the emergence of heterogeneous states.
Moreover, for $N\to\infty$ and a value of $\kappa$ as low as 4, a pattern formation requires $\mu<0.5$,
indicating that cross-inhibition has to be larger than self-inhibition.
This indicates that instabilities in undirected inhibitory networks correspond
to a multi-population competition-type (winner-takes-all) bifurcation.

% % %%%%%%%%%%%%%%%%%%%%%%%%%%%%%%%%%%%%%%%%%%%%%%%%%%%%%%%%%%%%%%%%%%%%%%%%
\begin{figure*}[t]
  \centerline{\includegraphics[width=\textwidth]{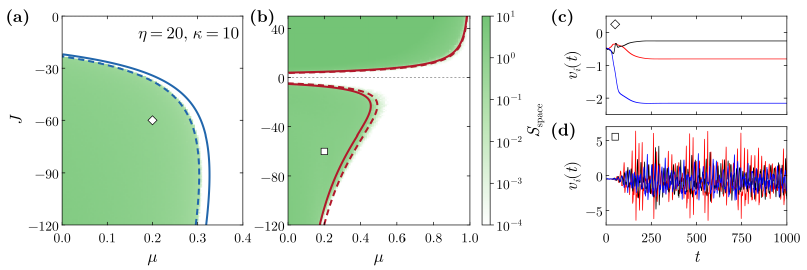}}
  \caption{Numerical validation of the bifurcation diagram.
        (a,b) Heatmaps of the spatial variability $S_\text{space}$ (Eq. \eqref{eq:S}) for $C$
        corresponding to an undirected network (panel (a)) and directed network (panel (b)) with $N=128$, $\kappa=10$, and $\eta = 20$.
        Blue and red solid curve are as in Fig. \ref{fig:bifurcationdiagram}(a). Blue and red dashed curve correspond
        to the stability boundaries given by Eq. \eqref{eq:msf} with the true values of $\Lambda_k$ obtained from the network.
        (c,d) Time series of variable $v_i$ for three randomly selected nodes corresponding to the undirected (panel (c)) and directed (panel (d)) networks for $(\mu,J)=(0.2,-60)$ (see $\Diamond$ and $\Box$ symbols in panels (a) and (b) respectively). 
        }
  \label{fig:validation}
\end{figure*}
% % %%%%%%%%%%%%%%%%%%%%%%%%%%%%%%%%%%%%%%%%%%%%%%%%%%%%%%%%%%%%%%%%%%%%%%%%

In order to numerically test these results, 
we simulate a network with $N=128$ nodes and average degree $\kappa = 10$ for fixed $\eta = 20$
and varying $J$ and $\mu$. Each simulation was initialized close to the homogeneous fixed point (see Appendix \ref{ap:simulations}
for full details). After a transient time, we monitor the spatial and temporal variability of the resulting dynamics
with the variances:
\begin{equation}
    \sigma_{\text{space}}^2(t)  =  \frac{1}{N}\sum_{i=1}^N (v_i(t)-\overline v(t))^2
\end{equation}
and
\begin{equation}
    \sigma_{\text{time},i}^2  =  \langle v_i(t)-\overline v(t))^2 \rangle,
\end{equation}
and then computing the time and space averages of their respective standard deviations:
\begin{equation}\label{eq:S}
    S_{\text{space}}  = \langle  \sigma_{\text{space}}(t) \rangle
    \quad\text{and}\quad 
    S_{\text{time}} =  \frac{1}{N}\sum_{i=1}^N \sigma_{\text{time},i}.
\end{equation}

Figure \ref{fig:validation}(a) shows the heatmap of the spatial variability $S_{\text{space}}$, together
with the bifurcation line corresponding to $N\to \infty$ (solid blue line), and the 
bifurcation obtained by using the maximal $\Lambda_k$ for $k >2$ of the network in Eq. \eqref{eq:msf} (dashed blue curve).
Heterogeneous states emerge in perfect agreement with the bifurcation obtained using the network spectra,
but with a small offset with respect to the asymptotic model (solid curve). This difference
can be traced back to the finite size error of the asymptotic relation given by Eq. \eqref{eq:spectra_undir}.
Remarkably, the dynamics in the region of instabilities are stationary everywhere (monitored with $S_\text{time}$, not shown).
Figure \ref{fig:validation}(c) shows the time traces of three randomly chosen nodes for $(\mu,J)=(0.2,-60)$ (see $\Diamond$ symbol in panel (a)).

\subsubsection{Directed networks}

In directed networks, most $\Lambda_k$ are pairs of complex conjugates.
% In fact, results from random matrix theory suggest that the number of real eigenvalues reduces to zero as $N$ increases \cite{}.
Writing $\Lambda_k = \Lambda_k^{(R)}+i\Lambda_k^{(I)}$ and using the algebraic expression for the
principal value of a square root of a complex number, Eq. \eqref{eq:msf} reads:
\begin{align}\label{eq:complex_lambda}
   \lambda_k^{\pm} = 2v_0 \pm &\sqrt{a+ \sqrt{a^2+b^2} } \\
    &\pm i\sign\left(\Lambda_k^{(I)}\right) \sqrt{-a + \sqrt{a^2+b^2} }
\end{align}
where
\begin{equation}
    a =  \tau r_0 J \Lambda_k^{(R)} - 2(\pi\tau r_0)^2\quad\text{ and }\quad b = \tau r_0 J \Lambda_k^{(I)}.
\end{equation}
Therefore, transverse instabilities may arise through complex conjugates of $\lambda^\pm_k$ crossing the imaginary axis.

Figures~\ref{fig:spectra}(c,d) show an example of this situation.
Panel (c) depicts the connectivity spectra in the complex plane (gray and red circles), with the bulk of the spectra following
relation \eqref{eq:spectra_dir}. Blue lines show the boundary delimited by
$\Real[\lambda_k] = 0$. Therefore, red circles correspond to those eigenmodes that destabilize the homogeneous state.
Panel (d) shows directly the corresponding spectra of the system Jacobian, $\lambda_k$, in complex plane. 
A cloud of complex conjugate eigenvalues cross the imaginary axis, indicating thus the onset of transverse instabilities is caused
by multiple oscillatory components.% This scenario seems closely related with that studied in \cite{muscinelli2019} for the emergence
%of resonant chaos. 

In order to obtain analytical boundaries for the bifurcation, we insert
$\Lambda_k \approx \mu+(1-\mu)e^{i\theta}\sqrt{\kappa^{-1}-N^{-1}}$ into Eq. \eqref{eq:complex_lambda}.
First, we look for the maxima of $\Real[\lambda_k^{\pm}]$ as a function of $\theta$. 
Next, we study when this maximum crosses the imaginary axis, $$\max_{\theta} \{\Real[\lambda_k^{\pm}]\} = 0.$$
These calculations, provided in Appendix \ref{ap:diagrams}, ultimately lead to the following 
condition for the stability boundary:
\begin{equation}\label{eq:boundary_complex} % CHECK!
(\kappa^{-1}-N^{-1}) (1-\mu)^2 (\tau r_0)^2 J^2 + \frac{2\Delta^2 J\mu}{\pi^2}-4\Delta^2\tau r_0  = 0.
\end{equation}
Combining this equation with \eqref{eq:fp_relation} and treating $r_0$ as a free parameter
gives the bifurcation diagrams of the system.

Red shaded regions in Figs. \ref{fig:bifurcationdiagram}(a,b) show the regions of transverse instabilities
in the $(J,\mu)$ and $(\eta,J)$ parameter spaces for $N\to\infty$.
The homogeneous state becomes unstable for both excitatory and inhibitory systems. 
In both cases the bifurcation exists for strong recurrent coupling (e.g., $\mu = 0.9$). 
For the case of excitatory units, the instabilities might occur for any $\eta$,
whereas inhibition requires a majority of neurons in a regular spiking regime ($\eta>0$).
Also, as it can be seen from Eq. \eqref{eq:msf}, these figures show that instabilities
might only arise if the underlying fixed point is a stable focus, as if $\lambda_1\in\mathbb{R}$, then $\Real[\lambda_k^+]\leq \lambda_1 < 1$
for all $k=2,\dots,N$.

Similarly to the undirected case, increasing network connectivity $\kappa$ cause
the region of non trivial dynamics to shrink (see salmon and pink curves in panel Figs. \ref{fig:bifurcationdiagram}(a)).
In fact, letting $\kappa \to N$ in Eq. \eqref{eq:boundary_complex} and keeping $J$ finite gives
$$J\mu = 2\pi^2\tau r_0.$$
However, if $(r_0,v_0)$ is a focus in Eq. \eqref{eq:homogeneous} then $J<2\pi^2\tau r_0$ (see Eq. \eqref{eq:appen_focus}),
thus the bifurcation condition cannot be fulfilled.
Therefore, instabilities in dense networks require $J\to \pm\infty$ 
to balance the first term in  Eq. \eqref{eq:boundary_complex} \footnote{If $\kappa \to N$, for $J>0$ one needs at least $J\approx (\kappa^{-1}-N^{-1})^{-1/4}$
since for large $J$, $J\approx \pi^2 \tau r_0$ (see Eq. \eqref{eq:fp_relation}).
For $J<0$, instabilities require, at least, $J\approx -(\kappa^{-1}-N^{-1})^{-3/2}$.}.

In order to test the bifurcation boundaries provided by our analysis, Fig. \ref{fig:validation}(b) shows the outcome of
simulations for $C$ obtained from a directed network with $N=128$ and $\kappa = 10$.
Sweeping the $(\mu,J)$ parameter space with fixed $\eta=20$ shows 
the emergence of heterogeneous dynamics in complete agreement  with the bifurcation boundary
obtained by a direct diagonalization of $C$ (dashed red curve), and with a small offset
with respect to the asymptotic case with $N\to\infty$ given by Eq. \eqref{eq:boundary_complex}.
In this case, contrary to the undirected scenario, the resulting dynamics are time-varying 
in the entire instability region (monitored with $S_\text{time}$, results not shown).
Exemplary time series
of this situation are depicted in Fig. \ref{fig:validation}(d), corresponding to $(\mu,J)=(0.2,-60)$
(see the $\Box$ symbol in panel (b)).
In the next section we characterize these emerging spatiotemporal dynamics in detail.

\section{Lyapunov exponent analysis}

\subsection{Emergence of high-dimensional chaos}
% % %%%%%%%%%%%%%%%%%%%%%%%%%%%%%%%%%%%%%%%%%%%%%%%%%%%%%%%%%%%%%%%%%%%%%%%%
\begin{figure}[t]
  \centerline{\includegraphics[width=0.5\textwidth]{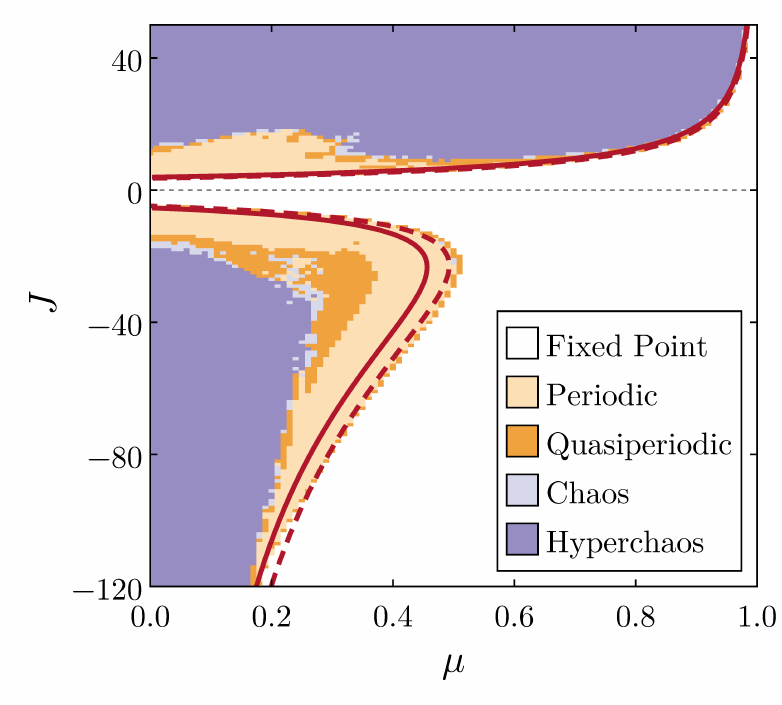}}
  \caption{Dynamical landscape in the $(\mu,J)$ parameter space for a directed network
  of $N=128$ nodes with $\kappa = 10$ and $\eta=20$. Colored region
  indicate the system dynamics, obtained from numerical computation of the two largest 
  Lyapunov exponents $\ell_1$ and $\ell_2$ (see text).
  Solid and dashed red curves as in Fig. \ref{fig:validation}.
  }
  \label{fig:le_diagram}
\end{figure}
% % %%%%%%%%%%%%%%%%%%%%%%%%%%%%%%%%%%%%%%%%%%%%%%%%%%%%%%%%%%%%%%%%%%%%%%%%

The previous analysis has uncovered Turing-like instabilities of system \eqref{eq:system1}
leading to heterogeneous fixed points in undirected networks
and spatiotemporal oscillatory states in directed networks which can be highly irregular (see Fig. \ref{fig:validation}(d)).
This section provides a detailed characterization of the dynamical complexity in the directed case,
mainly via numerical computation of the system's Lyapunov exponents $\ell_i$ for $i=1,\dots,2N$ \cite{Pikovsky2016}.

First, we compute the two largest Lyapunov exponents, $\ell_1$ and $\ell_2$ sweeping the $(\mu,J)$ parameter space.
This allows us to characterize the state of the system in numerical simulations with the following criteria:
\begin{itemize}
    \item Fixed point: $\ell_1<0$ and $\ell_2<0$.
    \item Limit-cycle: $\ell_1 = 0$ and $\ell_2<0$.
    \item Quasiperiodic torus: $\ell_1 = 0$ and $\ell_2 = 0$.
    \item Low-dimensional chaos: $\ell_1>0$ and $\ell_2 \leq 0$.
    \item High-dimensional chaos: $\ell_1>0$ and $\ell_2>0$.
\end{itemize}
In the simulations, we identify an exponent as zero if $|\ell_i|< 5\cdot 10^{-4}$.
Figure \ref{fig:le_diagram} shows these results for a network of $N=128$ populations with average connectivity $\kappa=10$ and fixed $\eta=20$. 
In the excitatory regime ($J>0)$ high-dimensional chaos dominates the region of instability,
with only a small island of periodic dynamics for low values of $\mu$. 
In the inhibitory case ($J<0$) high-dimensional chaos also prevails, but
always surrounded by a region of either periodic or quasiperiodic dynamics. 

\begin{figure*}[t]
  \centerline{\includegraphics[width=\textwidth]{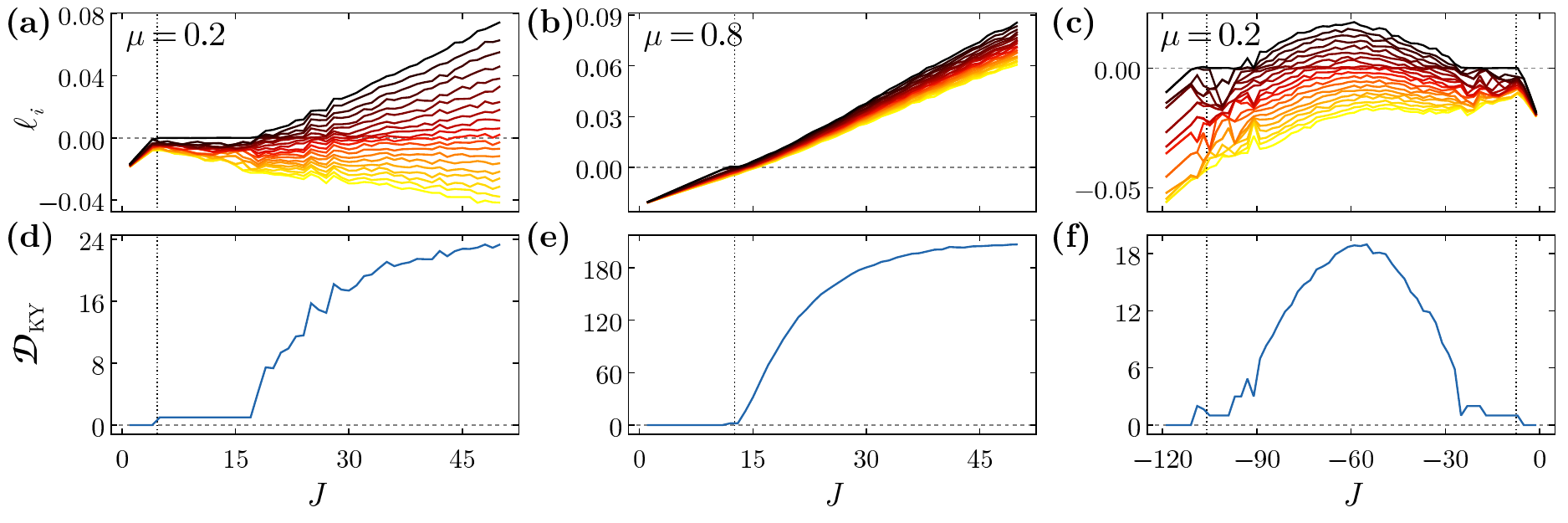}}
  \caption{Lyapunov analysis of \eqref{eq:system1} in directed networks. 
  (a,b,c) Twenty largest Lyapunov exponents computed numerically in a directed random network with $N=128$, $\kappa = 10$, and $\eta = 20$.
    Results correspond to excitatory network with $\mu=0.2$ (panel (a)) and $\mu=0.8$ (panel (b)), and an inhibitory network with $\mu=0.2$ (panel (c)).
  Notice the two separated ranges in the $y$-axis of panel (a), which we use to improve the visualization from periodic to chaotic dynamics.
  (d,e,f) Kaplan-Yorke fractal dimension computed from the full Lyapunov spectra, each panel corresponding to the same simulations as panels (a,b,c) respectively.
  Appendix \ref{ap:simulations} provides all the details on the numerical simulations.
  }
  \label{fig:lyapunov}
\end{figure*}

The next logical steps are to assess what transition leads to these high-dimensional chaotic states, and 
determining the fractal dimension of the chaotic attractor.
To do so, we compute the full Lyapunov spectrum for $\mu=0.2$ and $\mu=0.8$.
Figure \ref{fig:lyapunov}(a,b,c) shows the 20 largest Lyapunov exponents (out of 256 computed) for the same network as in Fig. \ref{fig:le_diagram}
as $J$ varies, while Fig. \ref{fig:lyapunov}(d,e,f) displays the corresponding Kaplan-Yorke fractal dimension \cite{Kaplan1979,Pikovsky2016}.

% % %%%%%%%%%%%%%%%%%%%%%%%%%%%%%%%%%%%%%%%%%%%%%%%%%%%%%%%%%%%%%%%%%%%%%%%%
% % %%%%%%%%%%%%%%%%%%%%%%%%%%%%%%%%%%%%%%%%%%%%%%%%%%%%%%%%%%%%%%%%%%%%%%%%

Panel (a) corresponds to a network of excitatory units for $\mu=0.2$, thus passing through the region of
periodic dynamics uncovered in Fig. \ref{fig:le_diagram}. 
Around $J\approx 17$ the largest exponents become positive consecutively, thus
signaling the onset of high-dimensional chaos.
Further increase of $J$ leads to an increase of both, the number of positive exponents and their magnitude.
The Kaplan-Yorke dimension (see Fig. \ref{fig:lyapunov}(d)) captures this behavior by showing the fractal dimension of the chaotic dynamics
increases regularly up to $\mathcal{D}_\text{KY}\approx 24$ for $J\approx 50$.

Panel (b) of Fig. \ref{fig:lyapunov} corresponds to the same excitatory network with increased recurrent coupling ($\mu=0.8$).
In this case, the transition to high-dimensional chaos is far more abrupt, with many exponents
becoming positive almost simultaneously just after the instability (indicated by the vertical black dotted line).
Here, the magnitude of the positive exponents remains comparable to that in the previous case ($\ell_i < 0.1$ in the range explored).
Nonetheless, the Kaplan-Yorke dimension (see Fig. \ref{fig:lyapunov}(e)) shows a strikingly high number of unstable directions,
reaching up to $\mathcal{D}_\text{KY}\approx 200$ in a system with 256 degrees of freedom.

Panel (c) of Fig. \ref{fig:lyapunov} corresponds to the inhibitory case with $\mu=0.2$.
Here, the chaotic region is bounded by states of periodic spatiotemporal activity ($\KY = 1$ in Fig. \ref{fig:lyapunov}(f)).
Furthermore, at the chaotic region, only a low number of exponents become positive, with a fractal dimension below $20$ in the
explored range. Further reducing $\mu$ can increase the dimensionality of the attractor (for instance, $\KY \approx 40$ for $(\mu,J)=(0.1,-60)$, results not shown). 

In all cases the chaotic dynamics consists of several unstable directions. For lower $\mu$,
this transition seems more regular than for larger $\mu$, where many exponents turn positive at once.
Together with the insights from linear stability analysis, these results suggest this is a case
of resonant high-dimensional chaos \cite{muscinelli2019}.

\subsection{Power spectra}

\begin{figure}[t]
  \centerline{\includegraphics[width=0.45\textwidth]{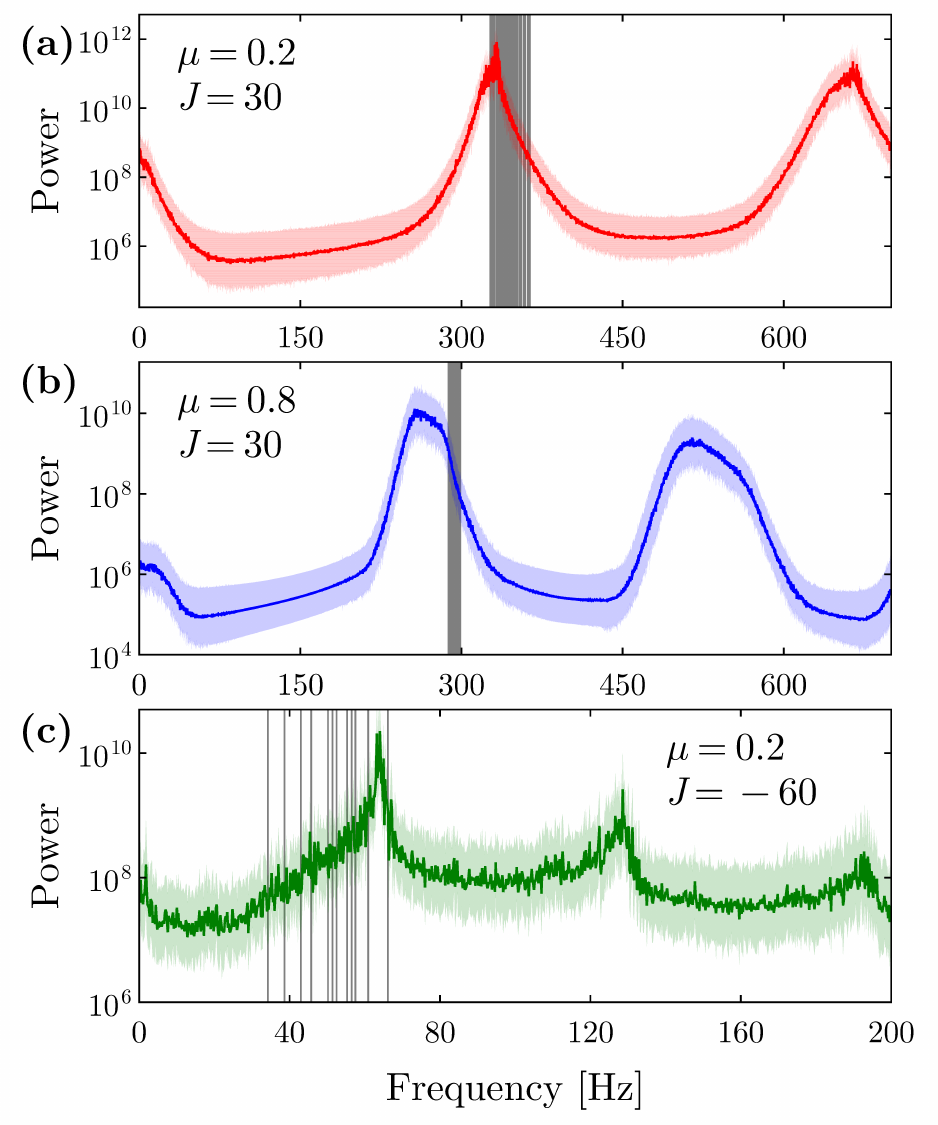}}
  \caption{(a-c) Average power spectra of variables $v_i(t)$ for $i=1,\dots,N$ for a directed network with $N=128$, $\kappa =10$,
  and $\eta = 20$.
  Solid lines represent the geometric mean, while the shaded regions indicate the geometric standard deviation factor.
  Grey vertical lines correspond to the frequencies given by Eq. \eqref{eq:f}.
  % Results correspond to time series of $5000$ ms, computed after discarding a suitable transient.
  }
  \label{fig:powerspectra}
\end{figure}

The simulations shown so far confirm the emergence of high-dimensional chaos in system \eqref{eq:system1}
with sparse directed random connectivities.
The instabilities leading to these states emerge from a pairs of complex eigenvalues \eqref{eq:complex_lambda}
crossing the imaginary axis. Thus we expect that, at least close to the bifurcation, the chaotic states contain oscillatory components 
at certain frequencies given by the imaginary part of $\lambda^{\pm}_k$ as
\begin{equation}\label{eq:f}
    f = \frac{\Imag[\lambda_k^{+}]}{2\pi \tau} 10^3\quad \text{(Hz)}.
\end{equation}
Moreover, a key interest of neural mass models is to unveil the mechanisms for the emergence
of neural rhythms. Thus in Fig. \ref{fig:powerspectra} we show the average power spectra
obtained for simulations with three different parameter combinations.
Vertical gray lines denote the frequencies corresponding to unstable modes, computed from Eq. \eqref{eq:f}.

For the excitatory cases (Fig. \ref{fig:powerspectra}(a) and (b)) show 
a prominent and broad peak covering a wide range of very fast frequencies
(around $300-400$ Hz in panel (a) and $250-350$ in panel (b)).
A slow aperiodic modulation also manifests through a bump around $0$ Hz. 
Instead, the inhibitory network (Fig. \ref{fig:powerspectra}(c)) shows
a sharper peak concentrated around $70$ Hz, and a less prominent modulation.
In all cases, the frequencies displayed in the simulations are within the frequency range captured by 
linear stability analysis in Eq.~\eqref{eq:f} (see gray vertical lines). 

Although model \eqref{eq:system1} does not aim to capture any specific biological phenomena,
the emergence of frequencies in the gamma range for the inhibitory system reflect
the importance of inhibition to this aim \cite{Brunel1999,Tiesinga2009,BW12}.
On the other hand, the super-fast frequencies observed in the excitatory networks
are more surprising, and their biological interpretation more limited, as we discuss in section \ref{sec:discussion}.

% % %%%%%%%%%%%%%%%%%%%%%%%%%%%%%%%%%%%%%%%%%%%%%%%%%%%%%%%%%%%%%%%%%%%%%%%%
% % %%%%%%%%%%%%%%%%%%%%%%%%%%%%%%%%%%%%%%%%%%%%%%%%%%%%%%%%%%%%%%%%%%%%%%%%

\subsection{Extensivity of chaos}

The emergence of high-dimensional chaos in a interconnected system is a strong
signature of extensive chaos \cite{ruelle1982}.
This phenomena consists of chaotic states in which the dimensionality of the chaotic attractor
grows linearly with the system size, thus reflecting that instabilities are mainly rooted in the local interactions. 
Extensive chaos is a common property of systems with spatial interaction in a continuous media \cite{manneville1985liapounov,livi1986,keefe1989,ohern1996,egolf2000,xi2000,paul2007},
but also of networks of coupled units \cite{monteforte2010,palmigiano2022,engelken2023b,clark2023,floriach2025}.

\begin{figure*}[t]
  \centerline{\includegraphics[width=0.95\textwidth]{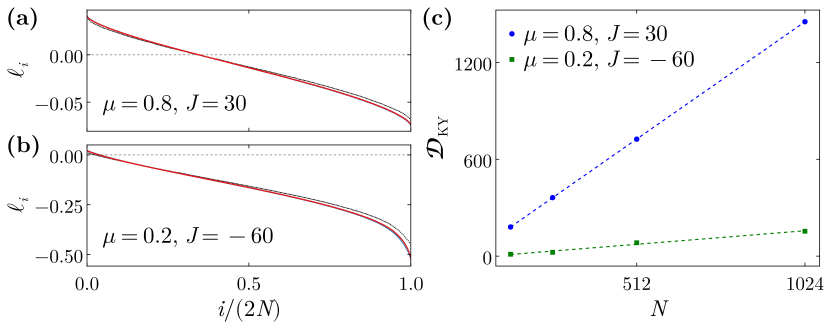}}
  \caption{Extensive chaos in system \eqref{eq:system1}. (a,b) Full Lyapunov spectra
  for $N=128$ (black), $N=512$ (blue), and $N=1024$ (red) for $(\mu,J)=(0.8,30)$ (panel (a)),
  and $(\mu,J)=(0.2,-60)$ (panel (b)). (c) Kaplan-Yorke dimension
 for $(\mu,J)=(0.8,30)$ (blue circles) and $(0.2,-60)$ (green squares). Dashed lines
 correspond to a linear regression. 
  }
  \label{fig:extensive}
\end{figure*}

Figures \ref{fig:extensive}(a,b) show the full Lyapunov spectra computed for $N=128,512$, and $1024$ with the same average local connectivity $\kappa = 10$ for $(\mu,J)=(0.2,-60)$ and $(\mu,J)=(0.8,30)$ \footnote{Results for $(\mu,J)=(0.2,30)$ where also computed, but are not shown,
as they are similar to those of $(\mu,J)=(0.2,-60)$.}. 
The indices $i=1,\dots,2N$ are normalized by the total number of degrees of freedom
to test for the existence of a well-defined thermodynamic limit.
For $N=128$ (red dots), the spectra closely matches that of $N=1024$, but with some appreciable
differences. However, differences between $N=512$ (blue dots) and $N=1024$ (black dots) are barely visible,
showing thus, a clear indication of extensive chaos.

In order to further provide numerical evidence of extensive chaos,
Figure  \ref{fig:extensive}(c) shows the scaling of the fractal dimension $\mathcal{D}_\text{KY}$
with the system size.
For all the parameter values tested,
the dimensionality of the chaotic attractor increases linearly
with $N$, with $\mu=0.2$ (green squares) displaying a lower slope than for the case with $\mu=0.8$ (blue circles).
This linear relation further confirms the extensivity of chaos in this system \footnote{
Notice that, to enforce a robust network model $C$ for $N\to\infty$, 
one should restrict the generation of Erdős-Rényi networks to strongly connected
topologies.
Otherwise, increasing $N$ with fixed $\kappa$, would surpass the percolation threshold, leading to the emergence of several connected components. 
Nonetheless, for $\kappa = 10$, networks with multiple components arise for $N\approx e^\kappa > 2.2\cdot 10^4$,
a system size well beyond our computational capabilities.}.

% % %%%%%%%%%%%%%%%%%%%%%%%%%%%%%%%%%%%%%%%%%%%%%%%%%%%%%%%%%%%%%%%%%%%%%%%%
% % %%%%%%%%%%%%%%%%%%%%%%%%%%%%%%%%%%%%%%%%%%%%%%%%%%%%%%%%%%%%%%%%%%%%%%%%

\section{Discussion}\label{sec:discussion}
 
In this work we studied networks of coupled next-generation neural mass models \cite{Montbrio2015}.
All nodes are described by the same identical dynamics capturing the behavior
of a population of QIF neurons. 
In isolation, each population only evolves towards steady states, corresponding to asynchronous activity of the underlying
neural network.
We have shown that random network interactions
lead to instabilitity of these states, allowing for the emergence of heterogeneous dynamics.

In the undirected case, instabilities only develop in sparse inhibitory networks
with a higher cross-coupling than self-feedback ($\mu<0.5$). 
The resulting states correspond to heterogeneous fixed points, corresponding
to a winner-takes-all situation arising from a mechanism analogous to Turing patterns in networked systems \cite{Nakao2010}.

By contrast, in directed networks, instabilities develop for both excitatory and inhibitory systems
and a wide range of parameter values, including large $\mu$.
Here sparsity is not a necessary condition, but it facilitates the emergence of non-trivial states (see Fig. \ref{fig:bifurcationdiagram}(a)). 
In this case the dynamics are time varying, with high-dimensional extensive chaos being the most prominent
state, as revealed by numerical simulations.  
These rhythmic irregular states are also analogous to the "topology driven" regimes arising from Turing bifurcations in directed networks \cite{Asllani2014}.

The analysis of system \eqref{eq:system1} requires the normalization condition \eqref{eq:normalization},
which ensures the existence of a homogeneous manifold with analytically treatable linear stability via the structural connectivity eigenvalues. 
This differs from reaction-diffusion systems \cite{Nakao2010,Asllani2014}, where diffusive coupling naturally allows for homogeneous states
and stability analysis requires the diagonalization of the network Laplacian
\footnote{In fact, the connectivity matrix $C$ is closely related to the normalized Laplacian of the network: $D^{-1}A$ is similar to $D^{-1/2}AD^{-1/2}$,
thus the spectra of $C$ and that of the normalized Laplacian $L = I_N+D^{-1/2}AD^{-1/2}$
coincide up to linear transformations involving $\mu$.}.
In spite of these differences, the mechanisms allowing for the emergence of non-trivial states in this work coincide with
that of the pattern formation formalism. 
Future work should assess to what extend the row-normalization influences
the bifurcation structure of the system and the emergence of chaotic states.

% Turing vs Sompolinsky
Apart from pattern formation, the emergence of complex states in networks of identical elements needs to be 
compared with another prominent conceptual framework: the seminal work from Sompolinsky, Crisanti and Sommers (SCS) \cite{sompolinsky1988}. 
This work, as well as subsequent studies \cite{kadmon2015a,crisanti2018,muscinelli2019,engelken2023b,pazo2024},
showed the emergence of high-dimensional extensive chaos arising in random neural networks through 
a homogeneous state suddenly destabilizing to a high number of eigenmodes crossing the imaginary axis.
In spite of the core similitude between our work and the SCS model, 
they differ on several technical aspects:

First, obviously, our model of choice is crucially different.
The SCS model is composed of simple rate neurons, which are equivalent to 
classical neural mass models \cite{castaldo2026}, with random connectivity.
Instead we use NG-NMMs, with single node dynamics given by a two-dimensional system. 
These changes impact the way in which instabilities occur:
In the SCS model the cloud of eigenvalues cross the imaginary as a disk centered in the real line.
Here, instead, the Jacobian eigenvalues in directed networks spread in two clouds of complex conjugate eigenvalues (see Fig. \ref{fig:spectra}(d)).
Our situation thus corresponds to an instance of resonant chaos, first unveiled in a variant of the SCS model with adaptation dynamics \cite{muscinelli2019}.

Second, our study clearly distinguishes the effects of excitation and inhibition in isolation,
allowing us to unveil the different role of the coupling type on the overall network dynamics.
This differs from SCS and related models, where neurons (or neural masses) may behave as a stimulator or a depressor depending
on the postsynaptic neuron.
This assumption breaks Dale's principle according to which neurons release the same neurotransmitters to all their postsynaptic cells \cite{strata1999}.
Our results thus show that excitation-inhibition interactions are not necessary to attain complex states in neural networks.

And the third main difference consists of the network structure, which here is binary and, usually sparse,
in opposition to the dense but random neural networks used in SCS. 
Therefore, our setup includes two additional topological parameters, the self coupling $\mu$, and
the network average degree $\kappa$, which prove key in determining the system stability.
Recent work has also covered this gap in the SCS by extending the dynamical mean field theory 
to sparse connectivities \cite{metz2025a}.

% Biological significance
We devote the rest of this discussion on the biological significance of our results.
Although the main motivation for this work has been theoretical, our findings in 
inhibitory networks have clear links to other modeling and experimental works.
For instance, a famous model for perceptual decision making consists of two cross-inhibiting populations \cite{wong2006,wong2007},
resulting into a winner-takes-all dynamics. In its simplest form, this model consists of two self-exciting and cross-inhibiting population,
giving rise to a pitchfork bifurcation. Our setup in undirected networks where $\mu<0.5$ closely matches this model,
with the external driving, mediated by $\eta$, as the excitation parameter. 

Additionally, the emergence of fast oscillatory dynamics in inhibitory directed networks is an instance of the well-known interneuron-network gamma (ING) mechanism \cite{Whittington2000,bartos2007,Tiesinga2009,BW12}.
This mechanism requires some form of synaptic kinetics or delay between synaptic interactions \cite{Brunel1999,brunel2003a,Devalle2017},
but it has also been shown to emerge in models with instantaneous interactions and sparse topologies \cite{diVolo2018,goldobin2024,goldobin2025}.
Here the situation compares to these latter cases, with the crucial differences that each node of our network represents
an entire population rather than a single neuron, and that the resulting dynamics here are, generally, chaotic.

In contrast, the super-fast rhythms unveiled in excitatory directed networks are more surprising and
difficult to interpret.
Excitation alone in neural systems is usually not enough to allow for the emergence of coordinated activity,
and the frequency ranges observed from our analysis are clearly beyond the typical ranges observed in mesoscopic recordings.
As a potential explanation, we remark that QIF neurons, from which Eqs. \eqref{eq:system1} are derived,
do not include refractory periods or ion channel recovery mechanisms. As a result, the firing rate grows unbounded with the input.
Real neurons, however, have depletion constraints that limit the firing frequency of neurons even for strong inputs.
NG-NMMs for QIF neurons with spike-frequency adaptation have been recently proposed \cite{ferrara2023,pietras2025}. 
These new models strongly modify the local dynamics of a single population, with the emergence of new oscillatory and even chaotic
regimes. 
Another interesting follow-up would be to understand how these super-fast rhythms modify in systems containing both, excitation and inhibition,
as it is the case in any biological neural setup.

\section*{Acknowledgements}
Work produced with the support of the grant PID2024-155942NB-I00 funded by MCIN/AEI/ 10.13039/501100011033 and ERDF, UE.
This work has also benefit from the UPC Dynamical Systems group’s cluster for research computing \url{https://dynamicalsystems.upc.edu/en/computing/}.
The author also thanks Ernest Montbrió, Diego Pazó, and Ivan León for helpful discussions.  

\section*{Data Availability}
The Julia codes used for simulation and analysis in this work are publicly available at \url{github.com/pclus/Coupled-NextGenNMM} \cite{repo}. 

\appendix

\section{Linear stability analysis}\label{ap:MSF}

Consider the tangent space dynamics of \eqref{eq:system1}, where the dynamics of the perturbation $(\delta r_i,\delta v_i)$
of a homogeneous fixed point $(r_i,v_i) = (r_0,v_0)$ $\forall i =1,\dots,N$ is given by:
\begin{equation}\label{eq:linear}
    \tau
    \begin{pmatrix}
        \dot \delta r_i\\ \dot \delta v_i
    \end{pmatrix}
    =
    \begin{pmatrix}
        2v_0 & 2 r_0\\
        -2(\pi \tau)^2 r_0 & 2v_0
    \end{pmatrix}
    +J\tau \sum_{j=1}^N c_{ij}
    \begin{pmatrix}
        0\\ \delta r_i
    \end{pmatrix}.
\end{equation}

For simplicity, let's define $\zeta_i = (\delta r_i,\delta v_i)^T$,
\begin{equation}
A=  \begin{pmatrix}
        2v_0 & 2 r_0\\
        -2(\pi \tau)^2 r_0 & 2v_0
    \end{pmatrix}
\quad
\text{and}
\quad
B= \begin{pmatrix}
    0 & 0\\
    J & 0
    \end{pmatrix}.
\end{equation}
With this notation, the linear system \eqref{eq:linear} reads
\begin{equation}\label{eq:linear2}
    \tau \dot \zeta_i = A \zeta_i + B\sum_{j=1}^N c_{ij} \zeta_i.
\end{equation}

Consider the diagonalization of $C$,
\begin{equation}\label{eq:eig}
 C \Psi_k = \Lambda_k \Psi_k
\end{equation}
where $\Lambda_k$ and $\Psi_k = (\Psi_{i k})_{i=1}^N\in \mathbb{C}^N$ are the eigenvalues and eigenvectors of $C$, with $k=1,\dots,N$.
We now express the perturbation $\zeta_k$ in the basis $\Psi_k$ as
\begin{equation}\label{eq:change}
\zeta_i(t) = \sum_{k=1}^N \Psi_{ik} \xi_k(t),
\end{equation}
where 
$\xi_k(t) \in\mathbb{C}^2$ are the coordinates of $\zeta_i$ in the eigenbasis $\{\Psi_1,\dots,\Psi_N\}$.

Inserting Eq.~\eqref{eq:change} into Eq.~\eqref{eq:linear2}
and making use of Eq.~\eqref{eq:eig} we have
\begin{equation}
\begin{aligned}
\tau \dot \zeta_i &= \tau \sum_{k=1}^N  \Psi_{ik} \dot \xi_k \\
&= A \sum_{k=1}^N \Psi_{ik} \xi_k + B \sum_{j=1}^N c_{ij}\sum_{k=1}^N \Psi_{jk} \xi_k \\
&= A \sum_{k=1}^N \Psi_{ik} \xi_k + B  \sum_{k=1}^N  \left(\sum_{j=1}^N c_{ij} \Psi_{jk}\right) \xi_k \\
&=\sum_{k=1}^N \left( A + \Lambda_k B  \right) \Psi_{ik} \xi_k .
\end{aligned}
\end{equation}
Since the eigenvectors $\Psi_k$ are a basis of $\mathbb{R}^N$, 
linear independence provides
\begin{equation}\label{eq:msf_final}
\dot \xi_k =  \left( A + \Lambda_k B \right)  \xi_k\;.
\end{equation}
Therefore, we decomposed the $2N\times 2N$ linear system Eq.~\eqref{eq:linear} 
into $N$ 2-dimensional linear systems that depend on the eigenvalues $\Lambda_k$.

\subsection{Compact form}

The previous linear stability analysis can be expressed in a more compact form by using matrix notation and the Kronecker product $\otimes$.
Let's define the perturbation vector
\begin{equation}
\zeta = (\delta r_1,\delta v_1,\dots,\delta r_N,\delta v_N)^T = (\zeta_1,\dots,\zeta_N)^T\in\mathbb{R}^{2N},    
\end{equation}
The full set of coupled linear systems \eqref{eq:linear2} for $i=1,\dots,N$ now reads
\begin{equation}\label{eq:linear3}
    \tau \dot \zeta = (I_N \otimes A + C \otimes B) \zeta.
\end{equation}

Let $\Lambda = (\Lambda_i\delta_{ij})$ be the diagonal matrix of eigenvalues of $C$,
and $\Psi =  (\Psi_1| \dots|\Psi_N)$ the corresponding matrix of eigenvectors, thus
\begin{equation}\label{eq:eig2}
 C \Psi = \Lambda \Psi.
\end{equation}
Let $\xi\in\mathbb{C}^{2N}$ be the coordinates of the perturbation $\zeta$ in the basis given by $\Psi$, i.e.,
\begin{equation}\label{eq:change2}
\zeta(t) = (\Psi \otimes I_2) \xi(t).
\end{equation}

Inserting Eq.~\eqref{eq:change2} into by Eq.~\eqref{eq:linear2}
we get
\begin{equation}\label{eq:aux1}
\begin{aligned}
    \tau \dot \zeta &= \tau (\Psi \otimes I_2) \dot \xi\\
    &= (I_N\otimes A + C \otimes B) (\Psi \otimes I_2) \xi  \\
&=  [(I_N\otimes A )(\Psi \otimes I_2) + (C\otimes B)(\Psi \otimes I_2)] \xi.
\end{aligned}
\end{equation}
The first term commutes, as can be seen by using the mixed-product property of $\otimes$:
\begin{equation}
\begin{aligned}
    (I_N\otimes A )(\Psi \otimes I_2) &= (I_N \Psi)\otimes ( A I_2 ) \\
    &= (\Psi I_N )\otimes ( I_2 A  ) = (\Psi \otimes I_2)(I_N\otimes A).
\end{aligned}
\end{equation}
The second term can be simplified using Eq. \eqref{eq:eig2}:
\begin{equation}
\begin{aligned}
    (C\otimes B)(\Psi \otimes I_2) &= (C\Psi)\otimes(B I_2) \\
    &= (\Lambda \Psi)\otimes(B I_2) \\
    &= (\Psi \Lambda )\otimes(I_2 B) 
    = (\Psi \otimes I_2 ) (\Lambda \otimes B). \\
\end{aligned}
\end{equation}
Therefore, Eq. \eqref{eq:aux1} reads
\begin{equation}
\begin{aligned}
    \tau (\Psi \otimes I_2) \dot \xi
    &= \left[(\Psi \otimes I_2)(I_N\otimes A)+(\Psi \otimes I_2 ) (\Lambda \otimes B)\right] \xi\\
    &=(\Psi \otimes I_2 ) \left( I_N\otimes A + \Lambda \otimes B \right) \xi
\end{aligned}
\end{equation}
Multiplying both sides by $\Psi^{-1} \otimes I_2$ we finally obtain
\begin{equation}
     \tau \dot \xi =  \left( I_N\otimes A + \Lambda \otimes B \right)\xi
\end{equation}
which is a block diagonal linear system equivalent to Eq. \eqref{eq:msf_final}.

\section{Bifurcation diagrams}\label{ap:diagrams}

Here we detail how to obtain the bifurcation diagrams depicted in Figures \ref{fig:homogeneous}
and \ref{fig:bifurcationdiagram}. 

\subsection{Homogeneous system}

The stability of homogeneous fixed point to homogeneous perturbations is given by $\Lambda_1=1$ in Eq. \eqref{eq:msf}.
Setting $\Real[\lambda^{\pm}_1] = 0$ provides
$$ J = \frac{\Delta^2}{2\pi^2(\tau r_0)^3}+2\pi^2(\tau r_0).$$
This equation can be interpreted as a parametric curve for $J$, with $r_0$ as a free parameter, $J=J(r_0)$.
Substituting this expression into \eqref{eq:fp_relation} and solving for $\eta$ provides:
$$ \eta(r_0) = \frac{-3\Delta^2}{(2\pi\tau r_0)^2}-(\pi \tau r_0)^2.$$
The graph of $(\eta(r_0),J(r_0))$ for $r_0>0$ provides the black curve in Fig. \ref{fig:homogeneous}.

The node-focus boundary can be obtained by imposing $\lambda_1^{+}-\lambda_1^{-} = 0$ in Eq. \eqref{eq:msf}, giving:
$$ J(r_0) = 2\pi^2 \tau r_0,$$
thus 
\begin{equation}\label{eq:appen_focus}
\eta(r_0) = \frac{-\Delta^2}{(2\pi\tau r_0)^2}-(\pi \tau r_0)^2.    
\end{equation}

\subsection{Transverse instabilities in undirected networks}

Solving Eq. \eqref{eq:transverse} for $\mu$ and Eq. \eqref{eq:fp_relation} for $J$ gives the parametric curves for the $(\mu,J)$ diagram (blue curve in Fig. \ref{fig:bifurcationdiagram}(a)):
\begin{align}
\mu(r_0) &= \frac{2 J(r_0) \sqrt{\kappa^{-1}-N^{-1}}-\frac{\Delta^2}{2\pi^2 (\tau r_0)^3} - 2\pi^2 \tau r_0}
{J(r_0)(1+2\sqrt{\kappa^{-1}-N^{-1}})},\\
J(r_0) & = \pi^2\tau r_0  - \frac{\eta}{\tau r_0} - \frac{\Delta^2}{4\pi^2(\tau r_0)^3}.
\end{align}

Similarly, solving Eq. \eqref{eq:transverse} for $J$ and Eq. \eqref{eq:fp_relation} for $\eta$
give the parametric curves for the $(\eta,J)$ diagram (blue curve in Fig. \ref{fig:bifurcationdiagram}(b)):
\begin{align}
J(r_0) &= -\frac{\frac{\Delta^2}{2\pi^2 (\tau r_0)^3} + 2\pi^2 \tau r_0}{
\mu-2(1-\mu)\sqrt{\kappa^{-1}-N^{-1}}},\\
\eta(r_0) &= -\frac{\Delta^2}{(2\pi\tau r_0)^2}+(\pi \tau r_0 )^2-J(r_0) \tau r_0.
 % (\pi \tau r_0 )^4-J\pi^2 (\tau r_0)^3 - \eta(\pi \tau r_0 )^2 = \frac{\Delta^2}{4}.
\end{align}

\subsection{Transverse instabilities in directed networks}
 
As indicated in the main text,
we substitute $\Lambda_k = \mu + (1-\mu)e^{i\theta}\sqrt{\kappa^{-1}-N^{-1}}$
in Eq. \eqref{eq:complex_lambda} and take the real part:
\begin{equation}\label{eq:AB}
\Real[\lambda_k^{\pm}]= 2v_0 + \sqrt{B+A\cos(\theta)+\sqrt{A^2+2AB\cos(\theta)+B^2}}    
\end{equation}
where
\begin{equation}
    \begin{aligned}
    v_0& = -\frac{\Delta}{2\pi \tau  r_0},\\
    A&=\sqrt{\kappa^{-1}}(1-\mu)\tau r_0 J,\\
    B&=\tau r_0 J \mu -2(\pi \tau r_0)^2.
    \end{aligned}
\end{equation}
We need to find the value of $\theta$ that maximizes Eq. \eqref{eq:AB}:
\begin{equation}
    \frac{d}{d\theta} \Real[\lambda_k^{\pm}] =0.
\end{equation}
Basic calculations provide the trivial critical values $\theta_1 =0$ and $\theta_2 = \pi$.
Additionally, if $2|B|>|A|$, we also obtain
\begin{equation}\label{eq:thetaopt}
    \theta_{3,4} = \pm\acos\left(\frac{-A}{2B}\right) 
    = \pm\acos\left(\frac{-\sqrt{\kappa^{-1}}(1-\mu) J}{2 J \mu -4\pi^2\tau r_0}\right). 
\end{equation}
The value of $\Real[\lambda^{\pm}]$ is the same for $\theta_3$ and $\theta_4$,
only the imaginary part changes sign.

Different arguments show that the bifurcation boundary corresponds to $\theta_{3,4}$. 
Indeed, $\theta_{1,2}$ correspond to real structural eigenvalues $\Lambda_k$. 
If $J>0$, we have from Eq. \eqref{eq:msf} that instabilities corresponding to real $\Lambda_k$
cannot occur for homogeneously stable fixed points (i.e., if $\Real[\lambda_1]<0$).
If $J<0$, substituting the expressions of $\theta_1$, $\theta_2$, and $\theta_{3,4}$ to Eq. \eqref{eq:AB}
shows that:
\begin{enumerate}
    \item $\theta_1=0$ corresponds to a minima of Eq. \eqref{eq:AB}, since vanishes the square root term. 
    \item If $\theta_{3,4}$ exist, they always correspond to eigenvalues with greatest or equal real part
    than those associated with $\theta_2=\pi$. 
    \item The eigenvalue associated to $\theta_2=\pi$ crossing the imaginary axis implies the existence of $\theta_{3,4}$.
\end{enumerate}

Therefore, in all cases the bifurcation boundary is obtained by substituting $\theta_{3,4}$
into $\Real[\lambda^{\pm}]=0$. After some manipulations, this provides
\begin{equation}
    A^2+8v_0^2 B =0
\end{equation}
or equivalently, Eq. \eqref{eq:boundary_complex}. 

To obtain the red, salmon and pink curves in Fig. \ref{fig:bifurcationdiagram}(a) we solve Eq. \eqref{eq:boundary_complex} for $\mu$ and \eqref{eq:fp_relation} for $J$ and plot the parametric curves $(\mu(r_0),J(r_0))$. Fig.  \ref{fig:bifurcationdiagram}(b)
instead if obtained by solving Eq. \eqref{eq:boundary_complex} for $J$
and \eqref{eq:fp_relation} for $\eta$. 
In either case, one has to solve a quadratic equation, which gives two solutions, corresponding to the excitatory and inhibitory cases respectively.

\section{Numerical simulations}\label{ap:simulations}

Numerical simulations have been performed using the DynamicalSystems.jl and ChaosTools.jl Julia packages  \cite{Datseris2018}. 
The code is publicly available in the Github repository \url{github.com/pclus/Coupled-NextGenNMM} \cite{repo}. 

All simulations use the Tsitouras 5/4 Runge-Kutta method with an adaptive time step $dt\leq 0.01$ and absolute and relative
tolerances of $10^{-6}$ and $10^{-5}$ respectively\cite{tsitouras2011runge}.
Computation of Lyapunov exponents use the common dynamical algorithm based on integration of the tangent space dynamics \cite{benettin1980,Pikovsky2016}.
The initial condition is set to the homogeneous fixed point of the system plus a random perturbation following a zero-mean normal distribution with $\sigma = 10^{-3}$.

Most simulations have a duration of $10^4$ time units after discarding $10^3$ of initial transient to the attractor,
and an additional $10^3$ of transient for the tangent space (thus $1.2\cdot 10^4$ time units in total).
The time window between renormalization (QR-decomposition calls) in the routines to compute Lyapunov exponents is set to $\Delta t = 1$.
We remark that these lengthy computations are required to achieve a good convergence close to the bifurcation boundaries.
For arbitrary parameter values, less integration time suffices to reach the asymptotic state.

The only exception to these computation parameters are the simulations presented in Fig. \ref{fig:extensive}.
In order to optimize computation resources via parallelization in a computing cluster, here exploited the ergodicity of chaotic systems
by reducing the duration of each simulation and performing averages across multiple realizations.
In particular,  computation of Lyapunov exponents consisted of $4\times 10^2$ time units,
performed after discarding an initial phase space transient of $10^3$ time units  with an additional $10^2$ time units of transient for the
tangent space.
For each parameter combination, these simulations were computed 20 times using different initial conditions.
The spectra and Kaplan-Yorke dimension presented in \ref{fig:extensive}
consist thus on the averages of these 20 simulations.
The corresponding standard deviations have been computed, but they are not shown since they are too small to be visibly appreciated in the figures.

\bibliography{references}

@article{Whittington2000,
title = {Inhibition-based rhythms: experimental and mathematical observations on network dynamics},
journal = {International Journal of Psychophysiology},
volume = {38},
number = {3},
pages = {315-336},
year = {2000},
issn = {0167-8760},
doi = {https://doi.org/10.1016/S0167-8760(00)00173-2},
url = {https://www.sciencedirect.com/science/article/pii/S0167876000001732},
author = {M.A Whittington and R.D Traub and N Kopell and B Ermentrout and E.H Buhl},
}

@Article{Bartos2007,
author={Bartos, Marlene
and Vida, Imre
and Jonas, Peter},
title={Synaptic mechanisms of synchronized gamma oscillations in inhibitory interneuron networks},
journal={Nature Reviews Neuroscience},
year={2007},
month={Jan},
day={01},
volume={8},
number={1},
pages={45-56},
issn={1471-0048},
doi={10.1038/nrn2044},
url={https://doi.org/10.1038/nrn2044}
}

@article{Tiesinga2009,
title = {Cortical Enlightenment: Are Attentional Gamma Oscillations Driven by ING or PING?},
journal = {Neuron},
volume = {63},
number = {6},
pages = {727-732},
year = {2009},
issn = {0896-6273},
author = {Paul Tiesinga and Terrence J. Sejnowski}
}

@Book{Pikovsky2016,
author={Pikovsky, Arkady
and Politi, Antonio},
title={Lyapunov Exponents: A Tool to Explore Complex Dynamics},
year={2016},
publisher={Cambridge University Press},
address={Cambridge},
isbn={9781107030428},
doi={10.1017/CBO9781139343473},
url={https://www.cambridge.org/core/books/lyapunov-exponents/0C96D0FCA7D41027741F485A9DA629AE},
url={https://doi.org/10.1017/CBO9781139343473}
}

@article{Nakao2010,
        author = {Nakao, Hiroya and Mikhailov, Alexander S.},
        doi = {10.1038/nphys1651},
        issn = {1745-2481},
        journal = {Nature Physics},
        number = {7},
        pages = {544-550},
        title = {Turing patterns in network-organized activator-inhibitor systems},
        url = {https://doi.org/10.1038/nphys1651},
        volume = {6},
        year = {2010}
}

@article{Asllani2014,
        author = {Asllani, Malbor and Challenger, Joseph D. and Pavone, Francesco Saverio and Sacconi, Leonardo and Fanelli, Duccio},
        doi = {10.1038/ncomms5517},
        issn = {2041-1723},
        journal = {Nature Communications},
        number = {1},
        pages = {4517},
        title = {The theory of pattern formation on directed networks},
        url = {https://doi.org/10.1038/ncomms5517},
        volume = {5},
        year = {2014}
}

@article{Pazo2016,
  title = {From Quasiperiodic Partial Synchronization to Collective Chaos in Populations of Inhibitory Neurons with Delay},
  author = {Paz\'o, Diego and Montbri\'o, Ernest},
  journal = {Phys. Rev. Lett.},
  volume = {116},
  issue = {23},
  pages = {238101},
  numpages = {5},
  year = {2016},
  month = {Jun},
  publisher = {American Physical Society},
  doi = {10.1103/PhysRevLett.116.238101}
}

@article{Montbrio2015,
  title = {Macroscopic Description for Networks of Spiking Neurons},
  author = {Montbri\'o, E. and Paz\'o, D. and Roxin, A.},
  journal = {Phys. Rev. X},
  volume = {5},
  issue = {2},
  pages = {021028},
  numpages = {15},
  year = {2015},
  publisher = {American Physical Society}
}

@article{Devalle2017,
    author = {Devalle, Federico AND Roxin, Alex AND Montbri\'o, Ernest},
    journal = {PLoS Computational Biology},
    publisher = {Public Library of Science},
    title = {Firing rate equations require a spike synchrony mechanism to correctly describe fast oscillations in inhibitory networks},
    year = {2017},
    month = {12},
    volume = {13},
    pages = {1-21},
    number = {12}
}

@article{Devalle2018,
  title = {Dynamics of a large system of spiking neurons with synaptic delay},
  author = {Devalle, Federico and Montbri\'o, Ernest and Paz\'o, Diego},
  journal = {Phys. Rev. E},
  volume = {98},
  issue = {4},
  pages = {042214},
  numpages = {13},
  year = {2018},
  month = {Oct},
  publisher = {American Physical Society},
  doi = {10.1103/PhysRevE.98.042214}
}

@article{Goldobin2021,
  title = {Reduction Methodology for Fluctuation Driven Population Dynamics},
  author = {Goldobin, Denis S. and di Volo, Matteo and Torcini, Alessandro},
  journal = {Phys. Rev. Lett.},
  volume = {127},
  issue = {3},
  pages = {038301},
  numpages = {6},
  year = {2021},
  month = {Jul},
  publisher = {American Physical Society}
}

@article{Ratas2019,
  title = {Noise-induced macroscopic oscillations in a network of synaptically coupled quadratic integrate-and-fire neurons},
  author = {Ratas, Irmantas and Pyragas, Kestutis},
  journal = {Phys. Rev. E},
  volume = {100},
  issue = {5},
  pages = {052211},
  numpages = {9},
  year = {2019},
  month = {Nov},
  publisher = {American Physical Society},
  doi = {10.1103/PhysRevE.100.052211},
  url = {https://link.aps.org/doi/10.1103/PhysRevE.100.052211}
}

@article{Pietras2019,
  title = {Exact firing rate model reveals the differential effects of chemical versus electrical synapses in spiking networks},
  author = {Pietras, Bastian and Devalle, Federico and Roxin, Alex and Daffertshofer, Andreas and Montbri\'o, Ernest},
  journal = {Phys. Rev. E},
  volume = {100},
  issue = {4},
  pages = {042412},
  numpages = {12},
  year = {2019},
  month = {Oct},
  publisher = {American Physical Society},
  doi = {10.1103/PhysRevE.100.042412},
  url = {https://link.aps.org/doi/10.1103/PhysRevE.100.042412}
}

@article{Dumont2019,
    doi = {10.1371/journal.pcbi.1007019},
    author = {Dumont, Grégory AND Gutkin, Boris},
    journal = {PLOS Computational Biology},
    publisher = {Public Library of Science},
    title = {Macroscopic phase resetting-curves determine oscillatory coherence and signal transfer in inter-coupled neural circuits},
    year = {2019},
    month = {05},
    volume = {15},
    url = {https://doi.org/10.1371/journal.pcbi.1007019},
    pages = {1-34},
    number = {5},

}

@incollection{Coombes2019,
  title={Next generation neural mass models},
  author={Coombes, Stephen and Byrne, {\'A}ine},
  booktitle={Nonlinear dynamics in computational neuroscience},
  pages={1--16},
  year={2019},
  publisher={Springer}
}

@article{Luke2013,
    author = {Luke, Tanushree B. and Barreto, Ernest and So, Paul},
    title = "{Complete Classification of the Macroscopic Behavior of a
                    Heterogeneous Network of Theta Neurons}",
    journal = {Neural Computation},
    volume = {25},
    number = {12},
    pages = {3207-3234},
    year = {2013},
    month = {12},
    issn = {0899-7667}
}

@article{Bi2020,
  title = {Coexistence of fast and slow gamma oscillations in one population of inhibitory spiking neurons},
  author = {Bi, Hongjie and Segneri, Marco and di Volo, Matteo and Torcini, Alessandro},
  journal = {Phys. Rev. Research},
  volume = {2},
  issue = {1},
  pages = {013042},
  numpages = {19},
  year = {2020},
  month = {Jan},
  publisher = {American Physical Society},
  doi = {10.1103/PhysRevResearch.2.013042},
  url = {https://link.aps.org/doi/10.1103/PhysRevResearch.2.013042}
}

@ARTICLE{Segneri2020,
AUTHOR={Segneri, Marco and Bi, Hongjie and Olmi, Simona and Torcini, Alessandro},
TITLE={Theta-Nested Gamma Oscillations in Next Generation Neural Mass Models},
JOURNAL={Frontiers in Computational Neuroscience},
VOLUME={14},
PAGES={47},
YEAR={2020}
}

@article{diVolo2018,
  title = {Transition from Asynchronous to Oscillatory Dynamics in Balanced Spiking Networks with Instantaneous Synapses},
  author = {di Volo, Matteo and Torcini, Alessandro},
  journal = {Phys. Rev. Lett.},
  volume = {121},
  issue = {12},
  pages = {128301},
  numpages = {6},
  year = {2018},
  month = {Sep},
  publisher = {American Physical Society},
  doi = {10.1103/PhysRevLett.121.128301},
  url = {https://link.aps.org/doi/10.1103/PhysRevLett.121.128301}
}

@article{Brunel1999,
    author = {Brunel, Nicolas and Hakim, Vincent},
    title = "{Fast Global Oscillations in Networks of Integrate-and-Fire Neurons with Low Firing Rates}",
    journal = {Neural Computation},
    volume = {11},
    number = {7},
    pages = {1621-1671},
    year = {1999},
    month = {10}
}

@article{BW12,
  title={Mechanisms of gamma oscillations},
  author={Buzs{\'a}ki, Gy{\"o}rgy and Wang, Xiao-Jing},
  journal={Annual Review of Neuroscience},
  volume={35},
  pages={203-225},
  year={2012},
  publisher={NIH Public Access}
}

@article{PP22,
  title = {Mean-field equations for neural populations with $q$-Gaussian heterogeneities},
  author = {Pyragas, Viktoras and Pyragas, Kestutis},
  journal = {Phys. Rev. E},
  volume = {105},
  issue = {4},
  pages = {044402},
  numpages = {10},
  year = {2022},
  month = {Apr},
  publisher = {American Physical Society}
}

@misc{Clusella2022,
  doi = {10.48550/ARXIV.2206.07521},
  
  url = {https://arxiv.org/abs/2206.07521},
  
  author = {Clusella, Pau and Köksal-Ersöz, Elif and Garcia-Ojalvo, Jordi and Ruffini, Giulio},
  
  title = {Comparison between an exact and a heuristic neural mass model with second order synapses},
  
  publisher = {arXiv},
  
  year = {2022},
  
  copyright = {Creative Commons Attribution 4.0 International}
}

@article{bordenave2012,
  title = {Circular Law Theorem for Random {{Markov}} Matrices},
  author = {Bordenave, Charles and Caputo, Pietro and Chafa{\"i}, Djalil},
  year = {2012},
  month = apr,
  journal = {Probability Theory and Related Fields},
  volume = {152},
  number = {3-4},
  pages = {751--779},
  issn = {0178-8051, 1432-2064},
  doi = {10.1007/s00440-010-0336-1},
  urldate = {2025-03-06},
  langid = {english}
}

@article{clusella_exact_2024,
    title = {Exact low-dimensional description for fast neural oscillations with low firing rates},
    volume = {109},
    issn = {2470-0045, 2470-0053},
    url = {https://link.aps.org/doi/10.1103/PhysRevE.109.014229},
    doi = {10.1103/PhysRevE.109.014229},
    number = {1},
    urldate = {2024-04-03},
    journal = {Physical Review E},
    author = {Clusella, Pau and Montbrió, Ernest},
    month = jan,
    year = {2024},
    pages = {014229},
}

@book{wilkinson1967,
  title={The Algebraic Eigenvalue Problem},
  author={Wilkinson, J.H.},
  series={Monographs on numerical analysis},
  year={1967},
  publisher={Clarendon Press}
}

@article{bick2020,
    title = {Understanding the dynamics of biological and neural oscillator networks through exact mean-field reductions: a review},
    volume = {10},
    issn = {2190-8567},
    shorttitle = {Understanding the dynamics of biological and neural oscillator networks through exact mean-field reductions},
    url = {https://mathematical-neuroscience.springeropen.com/articles/10.1186/s13408-020-00086-9},
    doi = {10.1186/s13408-020-00086-9},
    number = {1},
    urldate = {2022-02-03},
    journal = {The Journal of Mathematical Neuroscience},
    author = {Bick, Christian and Goodfellow, Marc and Laing, Carlo R. and Martens, Erik A.},
    month = dec,
    year = {2020},
    pages = {9},
}

@article{Datseris2018, doi = {10.21105/joss.00598}, url = {https://doi.org/10.21105/joss.00598}, year = {2018}, publisher = {The Open Journal}, volume = {3}, number = {23}, pages = {598}, author = {Datseris, George}, title = {DynamicalSystems.jl: A Julia software library for chaos and nonlinear dynamics}, journal = {Journal of Open Source Software} }

@article{tsitouras2011runge, title={Runge–Kutta pairs of order 5 (4) satisfying only the first column simplifying assumption}, author={Tsitouras, Ch}, journal={Computers \& Mathematics with Applications}, volume={62}, number={2}, pages={770–775}, year={2011}, publisher={Elsevier}, doi={10.1016/j.camwa.2011.06.002} }

@article{benettin1980,
  title = {Lyapunov {{Characteristic Exponents}} for Smooth Dynamical Systems and for Hamiltonian Systems; {{A}} Method for Computing All of Them. {{Part}} 2: {{Numerical}} Application},
  author = {Benettin, Giancarlo and Galgani, Luigi and Giorgilli, Antonio and Strelcyn, Jean-Marie},
  year = {1980},
  month = mar,
  journal = {Meccanica},
  volume = {15},
  number = {1},
  pages = {21--30},
  issn = {1572-9648},
  doi = {10.1007/BF02128237}
}

@misc{castaldo2026,
      title={Rosetta Stone of Neural Mass Models}, 
      author={Francesca Castaldo and Raul de Palma Aristides and Pau Clusella and Jordi Garcia-Ojalvo and Giulio Ruffini},
      year={2025},
      eprint={2512.10982},
      archivePrefix={arXiv},
      primaryClass={q-bio.NC},
      url={https://arxiv.org/abs/2512.10982}, 
}

@InProceedings{Kaplan1979,
author="Kaplan, James L.
and Yorke, James A.",
editor="Peitgen, Heinz-Otto
and Walther, Hans-Otto",
title="Chaotic behavior of multidimensional difference equations",
booktitle="Functional Differential Equations and Approximation of Fixed Points",
year="1979",
publisher="Springer Berlin Heidelberg",
address="Berlin, Heidelberg",
pages="204--227",
isbn="978-3-540-35129-0"
}

@techreport{reyner-parra2021,
    type = {preprint},
    title = {Phase-locking patterns underlying effective communication in exact firing rate models of neural networks},
    url = {http://biorxiv.org/lookup/doi/10.1101/2021.08.13.456218},
    doi = {10.1101/2021.08.13.456218},
    urldate = {2022-02-03},
    institution = {Neuroscience},
    author = {Reyner-Parra, David and Huguet, Gemma},
    month = aug,
    year = {2021},
}

@article{chen2022,
    title = {Exact mean-field models for spiking neural networks with adaptation},
    volume = {50},
    issn = {0929-5313, 1573-6873},
    url = {https://link.springer.com/10.1007/s10827-022-00825-9},
    doi = {10.1007/s10827-022-00825-9},
    number = {4},
    urldate = {2024-04-03},
    journal = {Journal of Computational Neuroscience},
    author = {Chen, Liang and Campbell, Sue Ann},
    month = nov,
    year = {2022},
    pages = {445--469},
}

@article{pietras2025,
    title = {Low-dimensional model for adaptive networks of spiking neurons},
    volume = {111},
    issn = {2470-0045, 2470-0053},
    url = {https://link.aps.org/doi/10.1103/PhysRevE.111.014422},
    doi = {10.1103/PhysRevE.111.014422},
    number = {1},
    urldate = {2026-03-18},
    journal = {Physical Review E},
    author = {Pietras, Bastian and Clusella, Pau and Montbrió, Ernest},
    month = jan,
    year = {2025},
    pages = {014422},
}

@article{pyragas2023,
    title = {Effect of {Cauchy} noise on a network of quadratic integrate-and-fire neurons with non-{Cauchy} heterogeneities},
    volume = {480},
    issn = {03759601},
    url = {https://linkinghub.elsevier.com/retrieve/pii/S0375960123003523},
    doi = {10.1016/j.physleta.2023.128972},
    urldate = {2023-10-17},
    journal = {Physics Letters A},
    author = {Pyragas, Viktoras and Pyragas, Kestutis},
    month = aug,
    year = {2023},
    pages = {128972},
}

@article{Gerster2021,
  title = {Patient-Specific Network Connectivity Combined With a Next Generation Neural Mass Model to Test Clinical Hypothesis of Seizure Propagation},
  volume = {15},
  ISSN = {1662-5137},
  url = {http://dx.doi.org/10.3389/fnsys.2021.675272},
  DOI = {10.3389/fnsys.2021.675272},
  journal = {Frontiers in Systems Neuroscience},
  publisher = {Frontiers Media SA},
  author = {Gerster,  Moritz and Taher,  Halgurd and {\v{S}}koch,  Anton{\'i}n and Hlinka,  Jaroslav and Guye,  Maxime and Bartolomei,  Fabrice and Jirsa,  Viktor and Zakharova,  Anna and Olmi,  Simona},
  year = {2021},
}

@article{Forrester2024,
  title = {Whole brain functional connectivity: Insights from next generation neural mass modelling incorporating electrical synapses},
  volume = {20},
  ISSN = {1553-7358},
  url = {http://dx.doi.org/10.1371/journal.pcbi.1012647},
  DOI = {10.1371/journal.pcbi.1012647},
  number = {12},
  journal = {PLOS Computational Biology},
  publisher = {Public Library of Science (PLoS)},
  author = {Forrester,  Michael and Petros,  Sammy and Cattell,  Oliver and Lai,  Yi Ming and O’Dea,  Reuben D. and Sotiropoulos,  Stamatios and Coombes,  Stephen},
  editor = {Gutkin,  Boris S.},
  year = {2024},
  pages = {e1012647}
}

@article{clusella2023,
  title = {Complex Spatiotemporal Oscillations Emerge from Transverse Instabilities in Large-Scale Brain Networks},
  author = {Clusella, Pau and Deco, Gustavo and Kringelbach, Morten L. and Ruffini, Giulio and {Garcia-Ojalvo}, Jordi},
  editor = {Gutkin, Boris S.},
  year = 2023,
  month = apr,
  journal = {PLOS Computational Biology},
  volume = {19},
  number = {4},
  pages = {e1010781},
  issn = {1553-7358},
  doi = {10.1371/journal.pcbi.1010781},
}

@article{Perl2023,
  title = {The impact of regional heterogeneity in whole-brain dynamics in the presence of oscillations},
  volume = {7},
  ISSN = {2472-1751},
  url = {http://dx.doi.org/10.1162/netn_a_00299},
  DOI = {10.1162/netn_a_00299},
  number = {2},
  journal = {Network Neuroscience},
  publisher = {MIT Press},
  author = {Perl,  Yonatan Sanz and Zamora-Lopez,  Gorka and Montbrió,  Ernest and Monge-Asensio,  Martí and Vohryzek,  Jakub and Fittipaldi,  Sol and Campo,  Cecilia González and Moguilner,  Sebastián and Ibañez,  Agustín and Tagliazucchi,  Enzo and Yeo,  B. T. Thomas and Kringelbach,  Morten L. and Deco,  Gustavo},
  year = {2023},
  pages = {632–660}
}

@article{pazo2025,
  title = {Low-Dimensional Dynamics of Globally Coupled Complex {{Riccati}} Equations: {{Exact}} Firing-Rate Equations for Spiking Neurons with Clustered Substructure},
  shorttitle = {Low-Dimensional Dynamics of Globally Coupled Complex {{Riccati}} Equations},
  author = {Paz{\'o}, Diego and Cestnik, Rok},
  year = 2025,
  month = may,
  journal = {Physical Review E},
  volume = {111},
  number = {5},
  pages = {L052201},
  issn = {2470-0045, 2470-0053},
  doi = {10.1103/PhysRevE.111.L052201},
  urldate = {2026-03-18},
  langid = {english}
}

@article{pietras2024,
  title = {Pulse {{Shape}} and {{Voltage-Dependent Synchronization}} in {{Spiking Neuron Networks}}},
  author = {Pietras, Bastian},
  year = 2024,
  month = jul,
  journal = {Neural Computation},
  volume = {36},
  number = {8},
  pages = {1476--1540},
  issn = {0899-7667, 1530-888X},
  doi = {10.1162/neco_a_01680},
  urldate = {2026-03-18},
  langid = {english}
}

@article{muscinelli2019,
    title = {How single neuron properties shape chaotic dynamics and signal transmission in random neural networks},
    volume = {15},
    issn = {1553-7358},
    url = {https://dx.plos.org/10.1371/journal.pcbi.1007122},
    doi = {10.1371/journal.pcbi.1007122},
    number = {6},
    urldate = {2025-03-19},
    journal = {PLOS Computational Biology},
    author = {Muscinelli, Samuel P. and Gerstner, Wulfram and Schwalger, Tilo},
    editor = {Latham, Peter E.},
    month = jun,
    year = {2019},
    pages = {e1007122},
}

@article{erdos1959,
    title = {On random graphs. {I}.},
    volume = {6},
    issn = {00333883},
    url = {https://publi.math.unideb.hu/load_doi.php?pdoi=10_5486_PMD_1959_6_3_4_12},
    doi = {10.5486/PMD.1959.6.3-4.12},
    number = {3-4},
    journal = {Publicationes Mathematicae Debrecen},
    author = {Erdős, P. and Rényi, A.},
    month = jul,
    year = {1959},
    pages = {290--297},
}

@misc{bordenave2010,
    title = {Spectrum of large random reversible {Markov} chains: two examples},
    shorttitle = {Spectrum of large random reversible {Markov} chains},
    url = {http://arxiv.org/abs/0811.1097},
    doi = {10.48550/arXiv.0811.1097},
    urldate = {2026-03-23},
    publisher = {arXiv},
    author = {Bordenave, Charles and Caputo, Pietro and Chafai, Djalil},
    month = may,
    year = {2010},
    note = {arXiv:0811.1097 [math]},
}

@book{bai2010,
    address = {New York, NY},
    series = {Springer {Series} in {Statistics}},
    title = {Spectral {Analysis} of {Large} {Dimensional} {Random} {Matrices}},
    copyright = {https://www.springernature.com/gp/researchers/text-and-data-mining},
    isbn = {978-1-4419-0660-1 978-1-4419-0661-8},
    url = {https://link.springer.com/10.1007/978-1-4419-0661-8},
    doi = {10.1007/978-1-4419-0661-8},
    language = {en},
    urldate = {2026-03-24},
    publisher = {Springer New York},
    author = {Bai, Zhidong and Silverstein, Jack W.},
    year = {2010},
}

@article{wigner1967,
    title = {Random {Matrices} in {Physics}},
    volume = {9},
    issn = {0036-1445, 1095-7200},
    url = {http://epubs.siam.org/doi/10.1137/1009001},
    doi = {10.1137/1009001},
    language = {en},
    number = {1},
    urldate = {2026-03-24},
    journal = {SIAM Review},
    author = {Wigner, Eugene P.},
    month = jan,
    year = {1967},
    pages = {1--23},
}

@article{girko1985,
    title = {Circular {Law}},
    volume = {29},
    issn = {0040-585X, 1095-7219},
    url = {http://epubs.siam.org/doi/10.1137/1129095},
    doi = {10.1137/1129095},
    language = {en},
    number = {4},
    urldate = {2026-03-24},
    journal = {Theory of Probability \& Its Applications},
    author = {Girko, V. L.},
    month = jan,
    year = {1985},
    pages = {694--706},
}

@article{wong2007,
  title = {Neural Circuit Dynamics Underlying Accumulation of Time-Varying Evidence during Perceptual Decision Making},
  author = {Wong, Kong-Fatt},
  date = {2007},
  journal = {Frontiers in Computational Neuroscience},
  shortjournal = {Front. Comput. Neurosci.},
  volume = {1},
  issn = {16625188},
  doi = {10.3389/neuro.10.006.2007},
  urldate = {2023-04-24},
  langid = {english}
}

@article{brunel2003a,
  title = {What {{Determines}} the {{Frequency}} of {{Fast Network Oscillations With Irregular Neural Discharges}}? {{I}}. {{Synaptic Dynamics}} and {{Excitation-Inhibition Balance}}},
  shorttitle = {What {{Determines}} the {{Frequency}} of {{Fast Network Oscillations With Irregular Neural Discharges}}?},
  author = {Brunel, Nicolas and Wang, Xiao-Jing},
  year = 2003,
  month = jul,
  journal = {Journal of Neurophysiology},
  volume = {90},
  number = {1},
  pages = {415--430},
  issn = {0022-3077, 1522-1598},
  doi = {10.1152/jn.01095.2002},
  urldate = {2022-02-15},
  langid = {english}
}

@article{ruelle1982,
  title = {Large volume limit of the distribution of characteristic exponents in turbulence},
  volume = {87},
  ISSN = {1432-0916},
  url = {http://dx.doi.org/10.1007/BF01218566},
  DOI = {10.1007/bf01218566},
  number = {2},
  journal = {Communications in Mathematical Physics},
  publisher = {Springer Science and Business Media LLC},
  author = {Ruelle,  David},
  year = {1982},
  month = jun,
  pages = {287–302}
}

@article{floriach2025,
    title = {From chimeras to extensive chaos in networks of heterogeneous {Kuramoto} oscillator populations},
    volume = {35},
    issn = {1054-1500, 1089-7682},
    url = {https://pubs.aip.org/cha/article/35/2/023115/3333443/From-chimeras-to-extensive-chaos-in-networks-of},
    doi = {10.1063/5.0243379},
    language = {en},
    number = {2},
    urldate = {2026-03-26},
    journal = {Chaos: An Interdisciplinary Journal of Nonlinear Science},
    author = {Floriach, Pol and Garcia-Ojalvo, Jordi and Clusella, Pau},
    month = feb,
    year = {2025},
    pages = {023115},
}

@article{engelken2023b,
    title = {Lyapunov spectra of chaotic recurrent neural networks},
    volume = {5},
    issn = {2643-1564},
    url = {https://link.aps.org/doi/10.1103/PhysRevResearch.5.043044},
    doi = {10.1103/PhysRevResearch.5.043044},
    language = {en},
    number = {4},
    urldate = {2026-03-26},
    journal = {Physical Review Research},
    author = {Engelken, Rainer and Wolf, Fred and Abbott, L. F.},
    month = oct,
    year = {2023},
    pages = {043044},
}

@article{wong2006,
    title = {A {Recurrent} {Network} {Mechanism} of {Time} {Integration} in {Perceptual} {Decisions}},
    volume = {26},
    issn = {0270-6474, 1529-2401},
    url = {https://www.jneurosci.org/lookup/doi/10.1523/JNEUROSCI.3733-05.2006},
    doi = {10.1523/JNEUROSCI.3733-05.2006},
    language = {en},
    number = {4},
    urldate = {2022-06-02},
    journal = {Journal of Neuroscience},
    author = {Wong, K.-F.},
    month = jan,
    year = {2006},
    pages = {1314--1328},
}

@article{goldobin2025,
  title = {Synaptic shot noise triggers fast and slow global oscillations in balanced neural networks},
  author = {Goldobin, Denis S. and Ageeva, Maria V. and di Volo, Matteo and Tixidre, Ferdinand and Torcini, Alessandro},
  journal = {Phys. Rev. E},
  volume = {112},
  issue = {3},
  pages = {034301},
  numpages = {23},
  year = {2025},
  month = {Sep},
  publisher = {American Physical Society},
  doi = {10.1103/47h5-fbyy},
  url = {https://link.aps.org/doi/10.1103/47h5-fbyy}
}

@article{goldobin2024,
  title = {Discrete Synaptic Events Induce Global Oscillations in Balanced Neural Networks},
  author = {Goldobin, Denis S. and di Volo, Matteo and Torcini, Alessandro},
  journal = {Phys. Rev. Lett.},
  volume = {133},
  issue = {23},
  pages = {238401},
  numpages = {6},
  year = {2024},
  month = {Dec},
  publisher = {American Physical Society},
  doi = {10.1103/PhysRevLett.133.238401},
  url = {https://link.aps.org/doi/10.1103/PhysRevLett.133.238401}
}

@article{ferrara2023,
  title = {Population spiking and bursting in next-generation neural masses with spike-frequency adaptation},
  author = {Ferrara, Alberto and Angulo-Garcia, David and Torcini, Alessandro and Olmi, Simona},
  journal = {Phys. Rev. E},
  volume = {107},
  issue = {2},
  pages = {024311},
  numpages = {16},
  year = {2023},
  month = {Feb},
  publisher = {American Physical Society},
  doi = {10.1103/PhysRevE.107.024311},
  url = {https://link.aps.org/doi/10.1103/PhysRevE.107.024311}
}

@inproceedings{manneville1985liapounov,
  title={Liapounov exponents for the Kuramoto-Sivashinsky model},
  author={Manneville, Paul},
  booktitle={Macroscopic Modelling of Turbulent Flows: Proceedings of a Workshop Held at INRIA, Sophia-Antipolis, France, December 10--14, 1984},
  pages={319--326},
  year={1985},
  organization={Springer}
}

@article{egolf2000,
  title = {Mechanisms of Extensive Spatiotemporal Chaos in {{Rayleigh}}--{{B{\'e}nard}} Convection},
  author = {Egolf, David A. and Melnikov, Ilarion V. and Pesch, Werner and Ecke, Robert E.},
  year = {2000},
  month = apr,
  journal = {Nature},
  volume = {404},
  number = {6779},
  pages = {733--736},
  issn = {0028-0836, 1476-4687},
  doi = {10.1038/35008013},
  copyright = {http://www.springer.com/tdm}
}

@article{keefe1989,
  title = {Properties of {{Ginzburg-Landau}} Attractors Associated with Their {{Lyapunov}} Vectors and Spectra},
  author = {Keefe, Laurence},
  year = {1989},
  month = oct,
  journal = {Physics Letters A},
  volume = {140},
  number = {6},
  pages = {317--322},
  issn = {03759601},
  doi = {10.1016/0375-9601(89)90627-0},
  copyright = {https://www.elsevier.com/tdm/userlicense/1.0/}
}

@article{livi1986,
  title = {Distribution of Characteristic Exponents in the Thermodynamic Limit},
  author = {Livi, R and Politi, A and Ruffo, S},
  year = {1986},
  month = aug,
  journal = {Journal of Physics A: Mathematical and General},
  volume = {19},
  number = {11},
  pages = {2033--2040},
  issn = {0305-4470, 1361-6447},
  doi = {10.1088/0305-4470/19/11/012}
}

@article{ohern1996,
  title = {Lyapunov Spectral Analysis of a Nonequilibrium {{Ising-like}} Transition},
  author = {O'Hern, Corey S. and Egolf, David A. and Greenside, Henry S.},
  year = {1996},
  month = apr,
  journal = {Physical Review E},
  volume = {53},
  number = {4},
  pages = {3374--3386},
  issn = {1063-651X, 1095-3787},
  doi = {10.1103/PhysRevE.53.3374},
  copyright = {http://link.aps.org/licenses/aps-default-license}
}

@article{paul2007,
  title = {Extensive chaos in Rayleigh-B\'enard convection},
  author = {Paul, M. R. and Einarsson, M. I. and Fischer, P. F. and Cross, M. C.},
  journal = {Phys. Rev. E},
  volume = {75},
  issue = {4},
  pages = {045203},
  numpages = {4},
  year = {2007},
  month = {Apr},
  publisher = {American Physical Society},
  doi = {10.1103/PhysRevE.75.045203},
  url = {https://link.aps.org/doi/10.1103/PhysRevE.75.045203}
}

@article{xi2000,
  title = {Extensive chaos in the Nikolaevskii model},
  author = {Xi, Hao-wen and Toral, Ra\'ul and Gunton, J. D. and Tribelsky, Michael I.},
  journal = {Phys. Rev. E},
  volume = {62},
  issue = {1},
  pages = {R17--R20},
  numpages = {0},
  year = {2000},
  month = {Jul},
  publisher = {American Physical Society},
  doi = {10.1103/PhysRevE.62.R17},
  url = {https://link.aps.org/doi/10.1103/PhysRevE.62.R17}
}

@article{palmigiano2022, title={Boosting of neural circuit chaos at the onset of collective oscillations}, url={https://doi.org/10.7554/eLife.90378.1}, DOI={https://doi.org/10.7554/eLife.90378.1}, journal = {eLife}, publisher={Cold Spring Harbor Laboratory}, author={Palmigiano, Agostina and Engelken, Rainer and Wolf, Fred}, year={2022}, month=aug }

@article{monteforte2010,
  title = {Dynamical Entropy Production in Spiking Neuron Networks in the Balanced State},
  author = {Monteforte, Michael and Wolf, Fred},
  journal = {Phys. Rev. Lett.},
  volume = {105},
  issue = {26},
  pages = {268104},
  numpages = {4},
  year = {2010},
  month = {Dec},
  publisher = {American Physical Society},
  doi = {10.1103/PhysRevLett.105.268104},
  url = {https://link.aps.org/doi/10.1103/PhysRevLett.105.268104}
}

@article{sompolinsky1988,
  title = {Chaos in {{Random Neural Networks}}},
  author = {Sompolinsky, H. and Crisanti, A. and Sommers, H. J.},
  date = {1988-07-18},
  journal = {Physical Review Letters},
  shortjournal = {Phys. Rev. Lett.},
  volume = {61},
  number = {3},
  pages = {259--262},
  issn = {0031-9007},
  doi = {10.1103/PhysRevLett.61.259},
  urldate = {2026-03-27},
  langid = {english}
}

@article{strata1999,
  title = {Dale's Principle},
  author = {Strata, Piergiorgio and Harvey, Robin},
  year = 1999,
  month = nov,
  journal = {Brain Research Bulletin},
  volume = {50},
  number = {5-6},
  pages = {349--350},
  issn = {03619230},
  doi = {10.1016/S0361-9230(99)00100-8},
  urldate = {2026-04-22},
  copyright = {https://www.elsevier.com/tdm/userlicense/1.0/}
}

@article{clark2023,
  title = {Dimension of {{Activity}} in {{Random Neural Networks}}},
  author = {Clark, David G. and Abbott, L. F. and Litwin-Kumar, Ashok},
  date = {2023-09-11},
  journal = {Physical Review Letters},
  shortjournal = {Phys. Rev. Lett.},
  volume = {131},
  number = {11},
  pages = {118401},
  issn = {0031-9007, 1079-7114},
  doi = {10.1103/PhysRevLett.131.118401},
  urldate = {2026-04-24},
  langid = {english}
}

@article{cestnik2026,
  title = {Two-Phase Quadratic Integrate-and-Fire Neurons: {{Exact}} Low-Dimensional Description for Ensembles of Finite-Voltage Neurons},
  shorttitle = {Two-Phase Quadratic Integrate-and-Fire Neurons},
  author = {Cestnik, Rok},
  year = 2026,
  month = mar,
  journal = {Physical Review Research},
  volume = {8},
  number = {1},
  pages = {L012049},
  issn = {2643-1564},
  doi = {10.1103/tq4x-8ny1},
  urldate = {2026-04-24},
  langid = {english}
}

@article{mayora-cebollero2025,
  title = {Dynamics of Coupled Neural Populations: {{The}} Role of Synaptic Dynamics},
  shorttitle = {Dynamics of Coupled Neural Populations},
  author = {{Mayora-Cebollero}, Ana and Barrio, Roberto and Li, Lei and {Mayora-Cebollero}, Carmen and P{\'e}rez, Luc{\'i}a},
  year = 2025,
  month = jun,
  journal = {Chaos: An Interdisciplinary Journal of Nonlinear Science},
  volume = {35},
  number = {6},
  pages = {063140},
  issn = {1054-1500, 1089-7682},
  doi = {10.1063/5.0219780},
  urldate = {2026-04-24},
  langid = {english}
}

@article{delicado-moll2026,
  title = {Emergent Spatiotemporal Dynamics in Large-Scale Brain Networks with next Generation Neural Mass Models},
  author = {{Delicado-Moll}, Rosa Maria and Huguet, Gemma and Clusella, Pau},
  year = 2026,
  month = sep,
  journal = {Physica D: Nonlinear Phenomena},
  volume = {493},
  pages = {135232},
  issn = {01672789},
  doi = {10.1016/j.physd.2026.135232},
  urldate = {2026-04-24},
  langid = {english}
}

@article{metz2025a,
  title = {Dynamical {{Mean-Field Theory}} of {{Complex Systems}} on {{Sparse Directed Networks}}},
  author = {Metz, Fernando L.},
  year = 2025,
  month = jan,
  journal = {Physical Review Letters},
  volume = {134},
  number = {3},
  pages = {037401},
  issn = {0031-9007, 1079-7114},
  doi = {10.1103/PhysRevLett.134.037401},
  urldate = {2026-05-11}
}

@software{repo,
  author       = {Pau Clusella},
  title        = {pclus/Coupled-NextGenNMM: First release - archived},
  month        = may,
  year         = 2026,
  publisher    = {Zenodo},
  version      = {v1.0-archive},
  doi          = {10.5281/zenodo.20180582},
  url          = {https://doi.org/10.5281/zenodo.20180582},
}

@article{kadmon2015a,
  title = {Transition to {{Chaos}} in {{Random Neuronal Networks}}},
  author = {Kadmon, Jonathan and Sompolinsky, Haim},
  year = 2015,
  month = nov,
  journal = {Physical Review X},
  volume = {5},
  number = {4},
  pages = {041030},
  issn = {2160-3308},
  doi = {10.1103/PhysRevX.5.041030},
  urldate = {2025-03-13},
  copyright = {http://creativecommons.org/licenses/by/3.0/},
  langid = {english}
}

@article{pazo2024,
  title = {Discontinuous Transition to Chaos in a Canonical Random Neural Network},
  author = {Paz{\'o}, Diego},
  year = 2024,
  month = jul,
  journal = {Physical Review E},
  volume = {110},
  number = {1},
  pages = {014201},
  issn = {2470-0045, 2470-0053},
  doi = {10.1103/PhysRevE.110.014201},
  urldate = {2025-03-06},
  langid = {english}
}

@article{crisanti2018,
  title = {Path Integral Approach to Random Neural Networks},
  author = {Crisanti, A. and Sompolinsky, H.},
  year = 2018,
  month = dec,
  journal = {Physical Review E},
  volume = {98},
  number = {6},
  pages = {062120},
  issn = {2470-0045, 2470-0053},
  doi = {10.1103/PhysRevE.98.062120},
  urldate = {2026-05-14},
  langid = {english}
}

\end{document}